\documentclass[iop]{emulateapj}
\usepackage{amsmath}
\usepackage{apjfonts}

\newcommand\xone{x_1}
\newcommand\xtwo{x_2}
\newcommand\xthree{x_3}
\newcommand\pc{\;{\rm pc}}
\newcommand\kpc{\;{\rm kpc}}
\newcommand\cs{c_s}

\newcommand\Msun{\; {\rm M}_{\odot}}
\newcommand\surf{\Msun\pc^{-2}}
\newcommand\kms{\; {\rm km}\;{\rm s}^{-1}}
\newcommand\freq{\kms\kpc^{-1}}

\newcommand\yr{\; {\rm yr}}
\newcommand\Myr{\;{\rm Myr}}
\newcommand\Gyr{\;{\rm Gyr}}
\newcommand\sfrunit{\;\Msun \kpc^{-2} \yr^{-1}}

\newcommand\Surf{\Msun\;{\rm pc^{-2}}}
\newcommand\dunit{\Msun\;{\rm kpc^{-3}}}
\newcommand\MBH{M_{\rm BH}}

\newcommand\RCR{R_{\rm CR}}

\newcommand\Omb{\Omega_{\rm b}}
\newcommand\Ombc{\Omega_{\rm b,crit}}
\newcommand\rhobar{\rho_{\rm bar}}

\newcommand\rhobul{\rho_{\rm bul}}
\newcommand\rhobulc{\rho_{\rm bul,crit}}

\newcommand\mring{f_{\rm ring}}

\newcommand\Rring{R_{\rm ring}}

\newcommand\Mbul{M_{\rm bul}}
\newcommand\C{{\mathcal C}}

\newcommand\Qb{Q_b}

\newcommand\ering{\epsilon_{\rm ring}}
\newcommand\simgt{\lower.5ex\hbox{$\; \buildrel > \over \sim \;$}}
\newcommand\simlt{\lower.5ex\hbox{$\; \buildrel < \over \sim \;$}}

\def\spose#1{\hbox to 0pt{#1\hss}}
\def\dt{\spose{\raise 1.0ex\hbox{\hskip2pt$\mathchar"201$}}}    

\shorttitle{Simulating Nuclear Rings}
\shortauthors{Li et al.}

\begin{document}

\title{Hydrodynamical Simulations of Nuclear Rings in Barred Galaxies} 

\author{Zhi Li\altaffilmark{1}, Juntai Shen\altaffilmark{1}, and Woong-Tae Kim\altaffilmark{2}}

\affil{$^1$Key Laboratory for Research in Galaxies and Cosmology, Shanghai Astronomical Observatory, Chinese Academy of Sciences, 80 Nandan Road, Shanghai 200030, China; jshen@shao.ac.cn \\
$^2$Center for the Exploration of the Origin of the Universe (CEOU), Astronomy Program, Department of Physics \& Astronomy, Seoul National University, Seoul 151-742,  Republic of Korea}



\begin{abstract}

Dust lanes, nuclear rings, and nuclear spirals are typical gas structures in the inner region of barred galaxies. Their shapes and properties are linked to the physical parameters of the host galaxy. We use high-resolution hydrodynamical simulations to study 2D gas flows in simple barred galaxy models. The nuclear rings formed in our simulations can be divided into two groups: one group is nearly round and the other is highly elongated. We find that roundish rings may not form when the bar pattern speed is too high or the bulge central density is too low. We also study the periodic orbits in our galaxy models, and find that the concept of inner Lindblad resonance (ILR) may be generalized by the extent of $\xtwo$ orbits. All roundish nuclear rings in our simulations settle in the range of $\xtwo$ orbits (or ILRs). However, knowing the resonances is insufficient to pin down the exact location of these nuclear rings. We suggest that the backbone of round nuclear rings is the $\xtwo$ orbital family, i.e. round nuclear rings are allowed only in the radial range of $\xtwo$ orbits. A round nuclear ring forms exactly at the radius where the residual angular momentum of infalling gas balances the centrifugal force, which can be described by a parameter $\mring$ measured from the rotation curve. The gravitational torque on gas in high pattern speed models is larger, leading to a smaller ring size than in the low pattern speed models. Our result may have important implications for using nuclear rings to measure the parameters of real barred galaxies with 2D gas kinematics.

\end{abstract}


\section{Introduction}
Stellar bars are the strongest disturber in the dynamical evolution of disk galaxies. They not only redistribute the stars but also have a significant impact on the interstellar medium (ISM) by introducing a non-axisymmetric gravitational torque inside the galaxy. Such a torque leads to the formation of interesting morphological substructures in the gaseous medium. Typical features include a pair of dust lanes related to shocks at the leading side of the bar (e.g., \citealt{ath92b}), a circum-nuclear star-forming ring (or nuclear ring for brevity) near the galactic center (e.g., \citealt{but86,gar91,bar95,mao01,maz08}), and nuclear spirals inside the ring which may be a channel for gas infall to fuel the central black hole (BH) (e.g., \citealt{shl90,reg99,kna00,dia03,lau04,van10,kim12a}). These substructures are commonly observed in nearby barred galaxies and are studied in previous simulations (e.g., \citealt {san76,rob79,sch81,van81,but96,mar03a,mar03b}). They may also be used as indicators to constrain galaxy properties (e.g., \citealt{wei99,wei01,per04}).

Nuclear rings are thought to form as a result of nonlinear interactions of gas with a bar potential (e.g., \citealt{shl90,hel94,pin95,kna95,but96,reg03,kim12a}). Due to the bar torque, gas readily forms dust-lane shocks at the leading side of the bar and flows inward along these dust lanes, forming a ring-like structure very close to the center. Observations show that usually nuclear rings are relatively circular in shape and have very high surface densities to trigger intense star formation (e.g., \citealt{bur60,kna06,maz08,com10,maz11}). The typical surface densities of star formation rate (SFR) in nuclear rings are of order $0.1$ to $1\sfrunit$, which is about two orders of magnitude larger than the disk-averaged SFR densities in normal galaxies. Therefore, previous studies suggested that starburst activities occurring in nuclear rings may help to build a pseudobulge, which is part of the secular evolution of the host galaxy (e.g., \citealt{kk04}).

While the formation mechanism of nuclear rings is still under debate, it has been widely believed that the location of nuclear rings is directly linked to the resonance radii. For example, \citet{com96} suggested that a nuclear ring forms near the inner Lindblad resonance (ILR) when there is only one ILR, as the torque outside (inside) ILR drives gas inflows (outflows), making it pile up at the ILR. Similarly, the nuclear ring forms between the inner ILR (iILR) and outer ILR (oILR) when there are two ILRs. This idea of resonance-driven ring formation requires that the bar torque dominates the thermal and ram pressures of the gas both inside and outside the ILR. There are also some studies which link nuclear rings to the periodic orbits in the bar potential. \citet{sell93} and \citet{sell14} suggested that the ring forms where orbits can be nested without intersecting each other. \citet{reg03,reg04} argued that the ring formation is more deeply related to the existence of $\xtwo$ orbits rather than the Lindblad resonances. However, the resolution of previous simulations are not high enough to study the details of gaseous structures in the central region of the host galaxy, and the spatial resolution is very important to simulate the gas flow in barred galaxies (\citealt{sor15}).

More recently, \citet{kim12a} revisited the issue of the formation of substructures in bars using high-resolution simulations by varying the gas sound speed $\cs$ and the mass of the central black hole ($\MBH$). \citet{kim12a} found that in models with smaller $\cs$, nuclear rings are narrower and located farther away from the center, while they become more spiral-like and boarder for larger $\cs$. \citet{kim12b} ran a number of numerical simulations by varying mass and axial ratio of the bar to study the effect of the bar strength on nuclear ring formation. They found that nuclear rings form due to the centrifugal barrier that the inflowing gas along the dust lanes experiences. They showed that the ring size is in general smaller than the outermost ILR radius and is smaller for a stronger bar due to greater loss of angular momentum at the dust lanes, consistent with the observational results of \citet{com10}. These results imply that the location of the nuclear ring is not only determined by the galactic gravitational field which generates various resonance radii and orbital families, but also affected by the properties of the interstellar medium itself such as the sound speed and magnetic field (e.g., \citealt{kul11,ks12}).

While the results of \citet{kim12b} are informative to understand why stronger bars host smaller nuclear rings, their models varied only the bar strength but fixed the other parameters such as the bar patten speed $\Omb$ and the axisymmetric part of the central galactic potential. However, $\Omb$ is one of the most important parameters of a bar. $\Omb$ is related to the total angular momentum content of the gas relative to the bar and is likely to control the strength and curvature of dust-lane shocks (e.g. \citealt{san15}). The central galactic potential is thought to provide the centrifugal barrier for the inflowing gas. Both of these two parameters are expected to affect the ring size. In this paper, we extend the models of \citet{kim12b} to explore the dependence of the ring size upon both $\Omb$ and the central galactic potential. To control the shape of the latter, we vary the bulge central density $\rho_{\rm bul}$ and the $M_{\rm BH}$. We measure the size, location, and thickness of the nuclear rings and study their relationships with the galactic parameters. These systematic experiments will offer more clues to the physical explanation of nuclear ring formation. We also compare our numerical results with observations of barred galaxies.

We note a few drawbacks of our numerical models from the outset. First of all, we consider an infinitesimally-thin 2D gaseous disk embedded in a rigid stellar gravitational potential, which ignores the potential effects of vertical gas motions introduced by finite disk thickness and back-reaction of the gas to the stellar components. Second, we adopt an isothermal equation of state, which simplifies what is physically a much more complex, turbulent  interstellar gas under radiative heating and cooling. Third, we neglect the gas self-gravity and star formation that occurs in high-density nuclear rings. As we discuss the caveats in Section~\ref{sec:limitations}, however, these approximations do not change our conclusions regarding the positions and shapes of nuclear rings significantly.

This paper is organized as follows. In Section 2, we describe our galaxy models, model parameters, and the numerical method. In Section 3, we present the simulation results for models with different $\Omb$, $\rhobul$ and $\MBH$. We discuss the formation of nuclear rings and compare our simulation results with observations in Section 4. In Section 5, we conclude with a short summary of our results.

\begin{figure}[!t]
\epsscale{1.1} \plotone{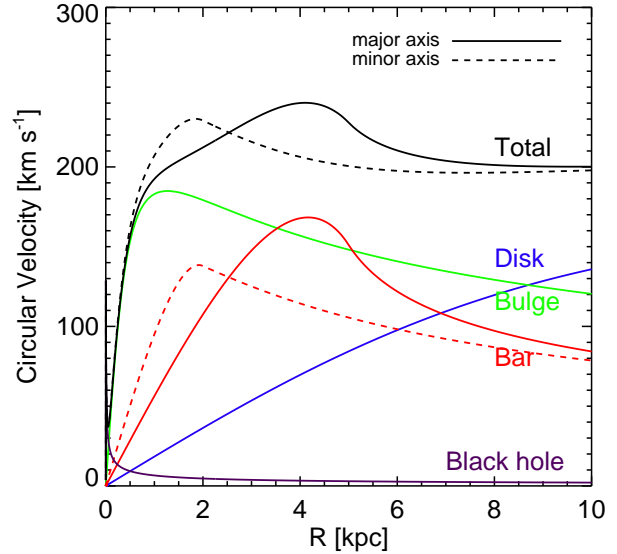}
\caption{Circular velocity of gas due to each galaxy component in model O33C24B10 along the bar major axis (solid) and minor axis (dashed). The black line is the total rotation curve and the other colors represent the contribution of different components. The BH affects the rotation curve only in the very central regions (with $R \simlt 0.1\kpc$).
\label{fig:rotcurve}}
\vspace{0.2cm}
\end{figure}

\section{Galaxy Models and Numerical Method}
\label{sec:modelsetup}

We study how gas respond to an imposed non-axisymmetric bar potential using hydrodynamic simulations. The bar is assumed to rotate rigidly about the galaxy center with a fixed pattern speed $\mathbf{\Omb}=\Omb\mathbf{\hat z}$. We solve the dynamical equations in the frame co-rotating with the bar in the $z=0$ plane. The equations of ideal hydrodynamics in this rotating frame are
\begin{equation}\label{eq:con}
\left(\frac{\partial}{\partial t}  + \mathbf{u}\cdot\nabla \right) \Sigma
= - \Sigma\nabla\cdot\mathbf{u},
\end{equation}
\begin{equation}\label{eq:mom}
\left(\frac{\partial}{\partial t}  + \mathbf{u}\cdot\nabla \right) \mathbf{u}
= -\cs^2 \frac{\nabla \Sigma}{\Sigma} - \nabla \Phi_{\rm ext} + \Omb^2
\mathbf{R} - 2\mathbf{\Omb}\times \mathbf{u}.
\end{equation}
Here $\Sigma$ and $\mathbf{u}$ denote the gas surface density and velocity, respectively. Equation \eqref{eq:con} is the well-known continuity equation and Equation \eqref{eq:mom} is the momentum equation in co-rotating cylindrical polar coordinates.

We use the grid-based MHD code \textit{Athena} \citep{gar05,sto08,pri08,sto09} in Cartesian geometry to integrate Equation \eqref{eq:con} and \eqref{eq:mom}. \textit{Athena} utilizes a higher-order Godunov scheme that conserves mass and momentum within machine precisions. It also provides several different schemes for integration in time, spatial reconstruction, and solution of the Riemann problem. We take the van Leer algorithm with piecewise linear reconstruction and first-order flux correction. For the Riemann solver, we use the exact nonlinear solver for our hydrodynamic models. Our simulation domain is a square box with size $L=20\kpc$ in each direction. We set up a uniform Cartesian grid with $4096\times4096$ cells over $|x|, |y|\leq L/2$. The corresponding grid spacing is $\Delta x=\Delta y=4.8\pc$. These high-resolution runs are necessary to explore the properties of bar substructures in detail (see also Section~\ref{sec:resolution}). Different from previous studies, we do not impose a point symmetry relative to the galaxy center, thus allowing odd-$m$ modes to grow in our models. We apply the outflow boundary conditions at the domain boundaries (i.e., at $|x|,|y|=L/2$). The outflow boundaries assume a zero gradient normal to the boundary for all flow variables except pressure. The solver extrapolates the required information from interior.

Equation \eqref{eq:mom} shows that there are three parameters that affect the simulation results. The first is the sound speed $\cs$ which effectively describes the thermal and turbulent properties of gas. The second parameter is the external galactic potential $\Phi_{\rm ext}$. A different shape of the potential could affect gas flows significantly. The third parameter is the angular speed of the bar $\Omb$, which is assumed to be constant over the radius. The setup of these three parameters is discussed in detail in the following.

\begin{deluxetable*}{lccccccccc}
\tabletypesize{\footnotesize} \tablewidth{0pt} \tablecaption{Model
Parameters and Nuclear Ring Radii
\label{tbl:model}} \tablehead{ \colhead{Model}   &
\colhead{$\Omb$}      & \colhead{$\rhobul$}      & \colhead{$\MBH$}      &
\colhead{$\RCR/a$}    & \colhead{Inner minor}    & \colhead{Outer minor} &
\colhead{Inner major} & \colhead{Outer major}    & \colhead{Ring type}
\\
\colhead{           } & \colhead{$\freq$} & \colhead{$10^{10}\dunit$} &
\colhead{$10^6\Msun$} & \colhead{       }  & \colhead{(pc)}            &
\colhead{(pc)}        & \colhead{(pc)}     & \colhead{(pc)}  & \colhead{  }
\\
\colhead{(1)} & \colhead{(2)} & \colhead{(3)} & \colhead{(4)} &
\colhead{(5)} & \colhead{(6)} & \colhead{(7)} & \colhead{(8)} &
\colhead{(9)} & \colhead{(10)}} \startdata
O21C24B10 & 21  &2.4 & 10 & 1.906 & 715.3 & 1169.4 & 695.8 & 979.0 &$\xtwo$\\
O23C24B10 & 23  &2.4 & 10 & 1.744 & 642.1 & 1115.7 & 676.3 & 920.4 &$\xtwo$\\
O25C24B10 & 25  &2.4 & 10 & 1.611 & 603.0 & 1032.7 & 676.3 & 920.4 &$\xtwo$\\
O27C24B10 & 27  &2.4 & 10 & 1.500 & 578.6 & 1018.1 & 578.6 & 871.6 &$\xtwo$\\
O29C24B10 & 29  &2.4 & 10 & 1.408 & 520.0 & 905.8  & 510.2 & 749.5 &$\xtwo$\\
O31C24B10 & 31  &2.4 & 10 & 1.330 & 485.8 & 812.9  & 461.4 & 690.9 &$\xtwo$\\
\bf{O33C24B10} & 33  &2.4 & 10 & 1.263 & 437.0 & 803.2  & 437.0 & 632.3 &$\xtwo$\\
O35C24B10 & 35  &2.4 & 10 & 1.205 & 393.1 & 744.6  & 388.2 & 573.7 &$\xtwo$\\
O37C24B10 & 37  &2.4 & 10 & 1.155 & 349.1 & 690.9  & 324.7 & 520.0 &$\xtwo$\\
O39C24B10 & 39  &2.4 & 10 & 1.111 & 329.6 & 617.7  & 339.3 & 476.1 &$\xtwo$\\
O41C24B10 & 41  &2.4 & 10 & 1.067 & 349.1 & 637.2  & 197.7 & 314.9 &$\xtwo$\\
O43C24B10 & 43  &2.4 & 10 & 1.040 & 173.3 & 270.9  & 407.7 & 642.1 &$\xone$\\
O45C24B10 & 45  &2.4 & 10 & 1.011 & 227.0 & 329.6  & 690.9 &1008.3 &$\xone$\\
O47C24B10 & 47  &2.4 & 10 & 0.984 & 129.3 & 197.7  & 622.5 &1140.1 &$\xone$\\
O49C24B10 & 49  &2.4 & 10 & 0.958 & 148.9 & 187.9  & 617.6 &1066.8 &$\xone$\\
\hline
\\
O33C12B10 & 33  &1.2 & 10 & 1.263 & 144.0 & 280.7  & 549.3 &1501.4 &$\xone$\\
O33C14B10 & 33  &1.4 & 10 & 1.263 & 119.6 & 275.8  & 646.9 &1652.8 &$\xone$\\
O33C16B10 & 33  &1.6 & 10 & 1.263 & 124.5 & 275.8  & 588.3 &1525.9 &$\xone$\\
O33C18B10 & 33  &1.8 & 10 & 1.263 & 280.7 & 681.1  & 148.9 & 397.9 &$\xtwo$\\
O33C20B10 & 33  &2.0 & 10 & 1.263 & 373.5 & 734.8  & 388.1 & 524.9 &$\xtwo$\\
O33C22B10 & 33  &2.2 & 10 & 1.263 & 388.1 & 788.5  & 397.9 & 568.8 &$\xtwo$\\
\bf{O33C24B10} & 33  &2.4 & 10 & 1.263 & 437.0 & 803.2  & 437.0 & 632.3 &$\xtwo$\\
O33C26B10 & 33  &2.6 & 10 & 1.263 & 495.6 & 837.4  & 466.3 & 656.7 &$\xtwo$\\
O33C28B10 & 33  &2.8 & 10 & 1.263 & 520.2 & 891.3  & 485.8 & 666.5 &$\xtwo$\\
O33C30B10 & 33  &3.0 & 10 & 1.263 & 544.4 & 905.7  & 520.0 & 705.6 &$\xtwo$\\
O33C32B10 & 33  &3.2 & 10 & 1.263 & 563.9 & 935.1  & 524.9 & 720.2 &$\xtwo$\\
O33C34B10 & 33  &3.4 & 10 & 1.263 & 588.4 & 939.9  & 539.5 & 725.1 &$\xtwo$\\
O33C36B10 & 33  &3.6 & 10 & 1.263 & 607.9 & 969.2  & 544.4 & 734.8 &$\xtwo$\\
O33C38B10 & 33  &3.8 & 10 & 1.263 & 622.5 & 969.2  & 573.7 & 734.8 &$\xtwo$\\
O33C40B10 & 33  &4.0 & 10 & 1.263 & 632.3 & 979.0  & 573.7 & 739.7 &$\xtwo$\\
\hline
\\
O25C16B10 & 25  &1.6 & 10 & 1.906 & 490.7 & 817.8  & 412.5 & 681.1 &$\xtwo$\\
O25C32B10 & 25  &3.2 & 10 & 1.906 & 866.7 & 1066.8 & 700.7 & 871.6 &$\xtwo$\\
O25C40B10 & 25  &4.0 & 10 & 1.906 & 905.7 & 1105.9 & 754.4 & 944.8 &$\xtwo$\\
O37C16B10 & 37  &1.6 & 10 & 1.155 & 183.1 & 305.1  & 568.8 & 979.0 &$\xone$\\
O37C32B10 & 37  &3.2 & 10 & 1.155 & 485.8 & 832.5  & 461.4 & 642.1 &$\xtwo$\\
O37C40B10 & 37  &4.0 & 10 & 1.155 & 563.9 & 905.7  & 529.7 & 676.2 &$\xtwo$\\
O43C16B10 & 43  &1.6 & 10 & 1.040 &$\cdots$&$\cdots$&$\cdots$&$\cdots$ &$\cdots$\\
O43C32B10 & 43  &3.2 & 10 & 1.040 & 378.4 & 642.1  & 378.4 & 520.0 &$\xtwo$\\
O43C40B10 & 43  &4.0 & 10 & 1.040 & 480.9 & 583.5  & 471.2 & 583.5 &$\xtwo$\\
\hline
\\
O45C24B100& 45  &2.4 & 100& 1.011 & 21.9  & 275.8  & 21.9  & 187.9 &$\xtwo$\\
O45C24B400& 45  &2.4 & 400& 1.011 & 266.1 & 549.3  & 231.9 & 427.2 &$\xtwo$

\enddata
\tablecomments{
Simulation parameters and the resulting nuclear ring radii. Col.(1): model ID. Col.(2): pattern speed of the bar. Col.(3): bulge central density. Col.(4): black hole mass. Col.(5): bar co-rotation radius over bar major axis. Col.(6)-Col.(9): the inner and outer nuclear ring radii along bar major and minor axes. Col.(10): nuclear ring type. The nuclear ring radii in the first two sets of this table is visualized in Figures \ref{fig:ringvsomegab} and \ref{fig:ringvscentralden}. Snapshots of gas surface density of the last two sets can be seen in Figures \ref{fig:snap} and \ref{fig:BHeffect}.}
\end{deluxetable*}

\begin{figure}[!t]
\epsscale{1.1} \plotone{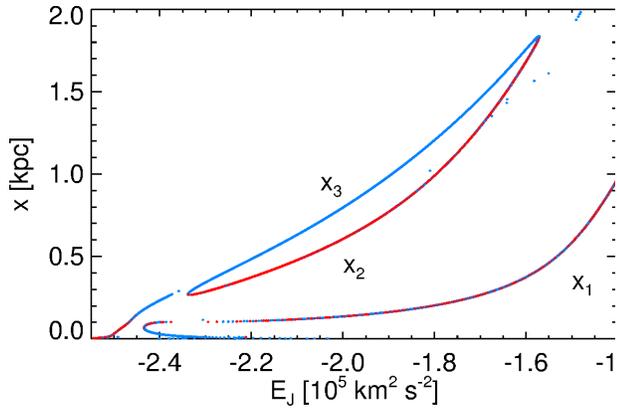}
\caption{Characteristic diagram or periodic orbital families for model O33C24B10. The intersection of each orbit with the $x$-axis (the bar minor axis) is plotted as a function of its Jacobi constant $E_J$. Each calculated periodic orbit is given by a point and we have not joined orbits of the same family by a continuous curve. Red and blue dots represent stable and unstable orbits. Note that the $\xtwo$ orbital family is stable and extends along the $x$-axis from $0.270$ to $1.835\kpc$.
\label{fig:orbitfamily}}
\vspace{0.2cm}
\end{figure}

The real ISM is multiphase and turbulent, with temperatures differing by as much as a few orders of magnitude (e.g., \citealt{fie69,mck77,mck07}). For simplicity, we treat the gaseous disk as being two-dimensional, isothermal, non-self-gravitating, and unmagnetized. The effects of $\cs$, self-gravity and magnetic fields are systematically explored in \citet{kim12a,kim12b} and \citet{ks12}. In this paper we fix $\cs$ to be $8\kms$ and set up an initially uniform gas disk with surface density $\Sigma_0=10\Surf$ at all radii, which are adopted from previous studies (e.g., \citealt{ath92b,kim12a}). Toomre's $Q$ for our gas disk is $\sim 2$ at all radii which means the disk is not strongly self-gravitating and is stable to the formation of spirals via gravitational instability.

The simulations setup for $\Phi_{\rm ext}$ is nearly identical to those in \citet{kim12a}. Here we briefly summarize our numerical models and highlight the differences between the current models and those of \citet{kim12a}. The gaseous disk is placed under a $\Phi_{\rm ext}$ consisting of four components: a Kuzmin-Toomre stellar disk (\citealt{kuz56}), a modified Hubble stellar bulge, a Ferrers ellipsoid stellar bar (\citealt{fer87,pfe84}), and a central BH represented by a Plummer sphere. We emphasize that a norminal dark matter halo component has not been included because the disk here acts as the dark matter halo to generate flat rotation curve for historical reasons (\citealt{ath92b}), so that we can compare our work directly with previous studies. These idealized models are designed to isolate out the effects of the $\Omb$, $\rhobul$ and $\MBH$ on the formation of nuclear rings.

To represent the elongated bar potential, we use an $n = 1$ inhomogeneous Ferrers prolate spheroid:
\begin{eqnarray}\label{eq:bar}
\rho =
\left\{ \begin{array}{ll}
\rhobar \left( 1-g^2 \right)^n & ~~~{\rm for}~ g<1, \\
~~~~~~~0 & ~~~{\rm elsewhere},
\end{array} \right.
\end{eqnarray}
\begin{equation}\label{eq:bar2}
g^2= y^2/a^2 +(x^2 +z^2)/b^2.
\end{equation}
The bar central density $\rhobar$, major axis $a$ and minor axis $b$ of the bar are fixed at $4.47 \times 10^{8}\dunit$, $5\kpc$, and $2\kpc$, respectively. This gives a bar axis ratio of $2.5$ and a bar mass of $1.5 \times 10^{10}\Msun$. The shape and mass of the bar are fixed for all of our models; their effects were investigated by \citet{kim12b}.

In addition to the bar, the bulge potential also affects the gas flow pattern and the ring size (e.g., \citealt{ath92b}). For the bulge potential, we take a modified Hubble profile:
\begin{equation}\label{eq:bulge}
\rho(r) = \rhobul \left(1 + \frac{r^2}{r_b^2}\right)^{-3/2},
\end{equation}
where $\rho_{\rm bul}$ and $r_b$ represent the bulge central density and scale length, respectively. \citet{maz11} argued that the ring size is well correlated with the compactness $\C\equiv V_0^2/R_{\rm to}$, where $R_{\rm to}$ is the turnover radius of the rotation curve that has a velocity $V_0$ for the flat(ter) part. Their results show that more compact (with smaller $R_{\rm to}$) galaxies have smaller rings (see Section~\ref{sec:compareobs} for more details). To test this idea, it is necessary to consider models with different compactness values, which we control by varying $\rhobul$.

Table \ref{tbl:model} lists the model parameters and some of the simulations results. Column (1) gives the model names. Columns (2)-(4) list model parameters, Columns (6)-(9) exhibit nuclear ring radii along bar major and minor axes, and Column (10) defines the nuclear ring type (see Section~\ref{sec:simulationresults}). The first set of models in Table \ref{tbl:model} explores the effects of $\Omb$, while the models in the second set adopt a different value of $\rhobul$. In the third set, we vary both $\Omb$ and $\rhobul$. The last set varies $\MBH$. Figure \ref{fig:rotcurve} plots the total circular rotation curves together with the contribution from each component for our canonical model O33C24B10 with $\Omb =33\freq$, $\rhobul = 2.4 \times 10^{10}\dunit$, and $\MBH = 1\times10^7\Msun$.

In order to avoid transients, we start with an axisymmetrized bar and the bar mass is linearly ramped up to its final maximum (\citealt{ath92b,kim12a}). The typical time scale for bar growth here (i.e. $T_{\rm grow} = 2\pi/\Omega_b$) is a few hundred$\Myr$. We have experimented with different bar growth times and confirmed that the gas distribution at the end of the runs is not sensitive to the bar growth times, although different bar growth times may cause distinct initial transient gas flow patterns, similar to the results in \citet{pat00}. All simulations are run for $5$ bar revolution times, by which time the gas distribution has reached a quasi-steady state in the bar co-rotation frame.

\begin{figure}[!t]
\epsscale{1.1} \plotone{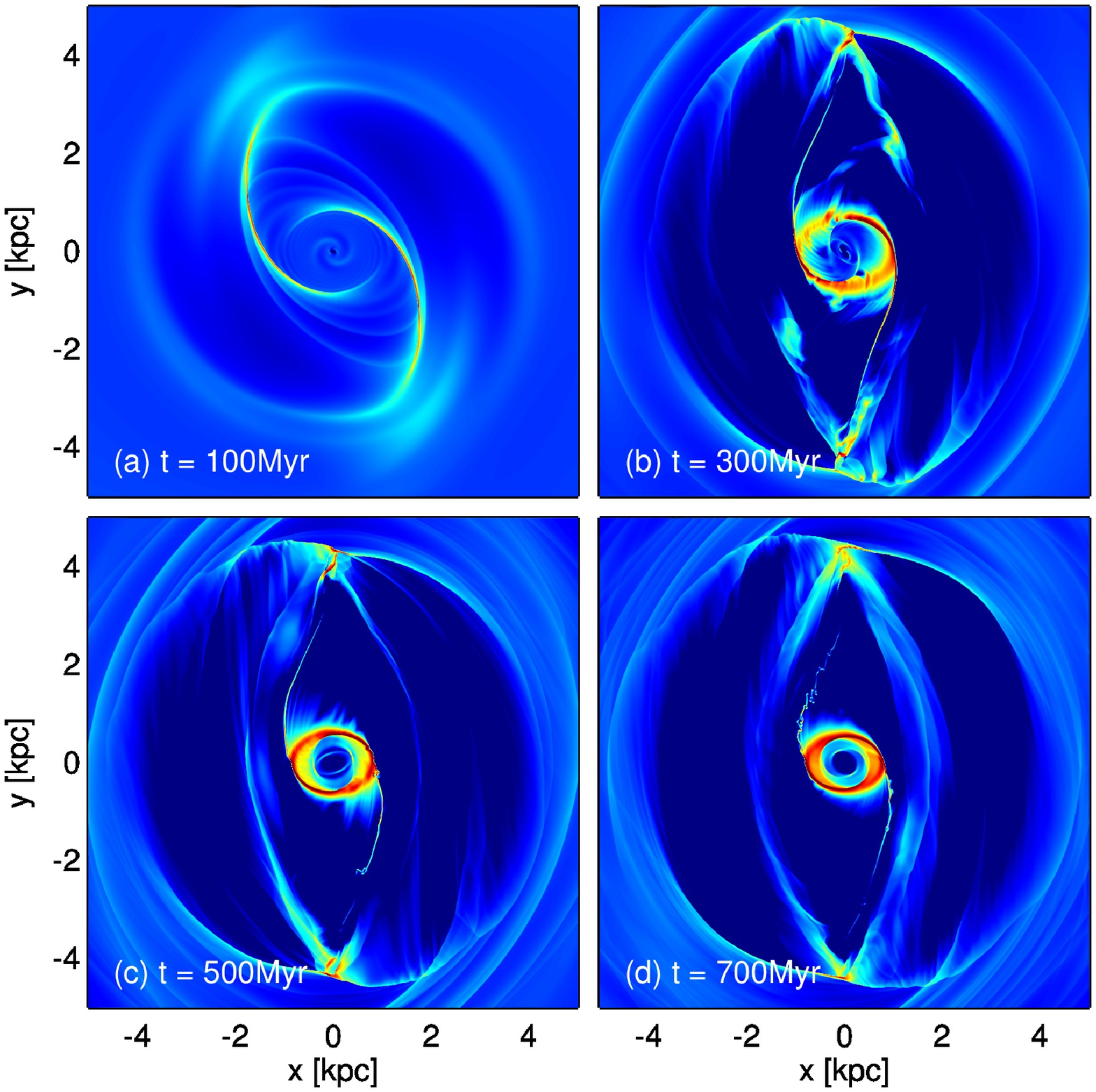}
\caption{Time evolution of the gas surface density in the bar co-rotating frame in Model O33C24B10. The snapshots are in logarithm scale and the bar is parallel to y-axis. Dust lanes (shocks), nuclear rings and nuclear spirals are clearly visible in these snapshots. (\textit{An animation of this figure is available.})
\label{fig:evo}}
\vspace{0.2cm}
\end{figure}

\begin{figure*}
\epsscale{1.0} \plotone{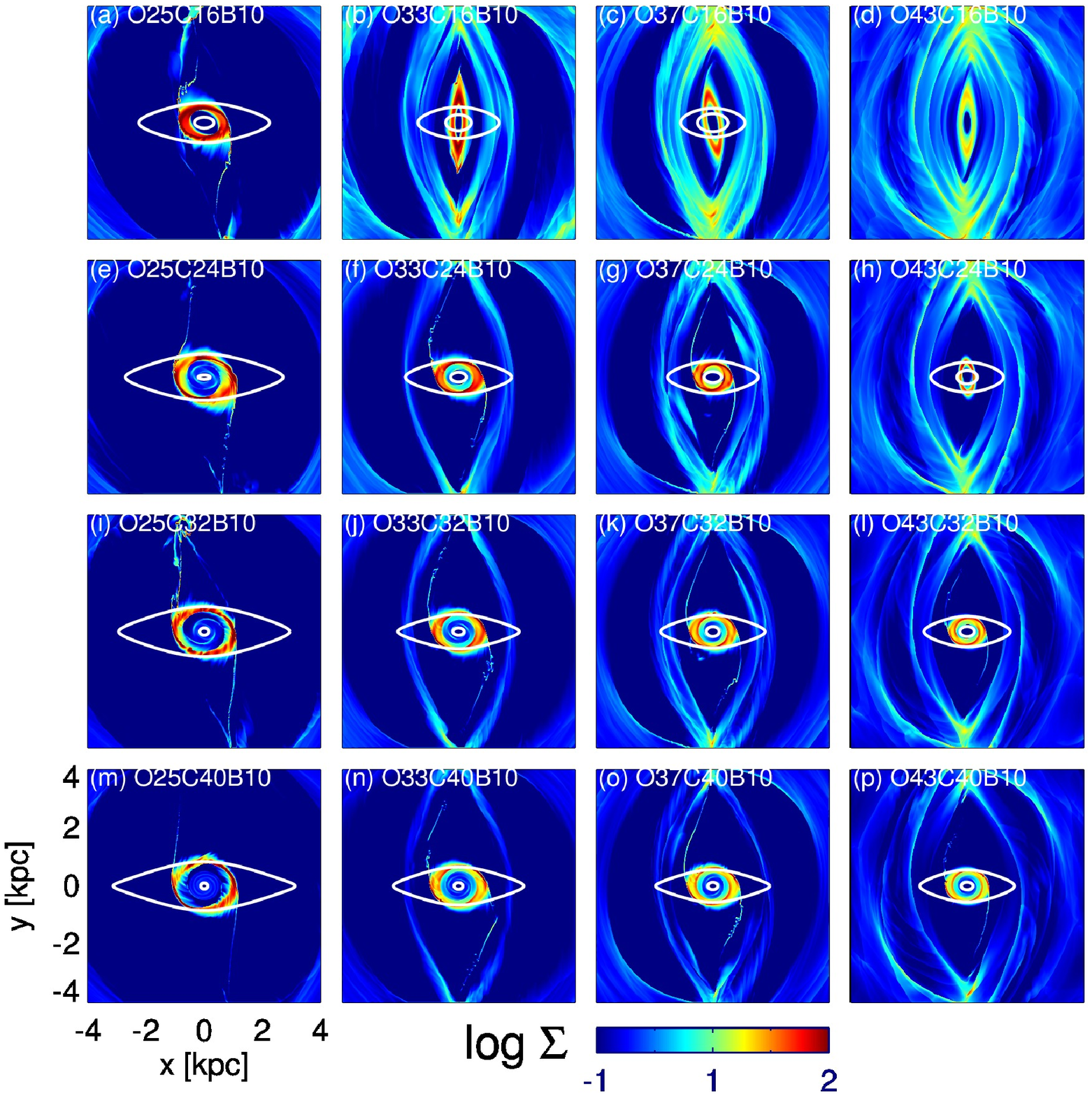} \caption{Gas surface density at $t = 800\Myr$ for the models. Each panel shows the inner $\pm4\kpc$ regions where the bar is oriented vertically along the $y$-axis. Each row corresponds to models with $\rhobul=1.6$, $2.4$, $3.2$, and $4.0 \times 10^{10}\dunit$ from top to bottom, while each column is for models with $\Omb=25$, $33$, $37$, and $43\freq$ from left to right. The white solid lines indicate the innermost and outermost $\xtwo$ orbits. Note that model O43C16B10 does not have $\xtwo$ orbits.
\label{fig:snap}}
\end{figure*}

\section{Simulation results}
\label{sec:simulationresults}

\subsection{Characteristic diagram of the potential}
We use characteristic diagrams to identify periodic orbital families in our galaxy models. The typical orbital families in a barred galaxy are $\xone$, $\xtwo$, and $\xthree$. The $\xone$ orbital family is the backbone of the bar, which is parallel to the bar major axis; $\xtwo$ and $\xthree$ are centrally concentrated and perpendicular to the bar major axis. Figure \ref{fig:orbitfamily} plots the orbital energy Jacobi constant $E_J$ vs. the distance $x$ where each orbit crosses the short axis of the bar. Each sequence of periodic orbital families generates a continuous curve, and each dot represents a periodic orbit. The $\xtwo$ and $\xthree$ orbital families form a banana shape in this characteristic curve at small $x$, the while $\xone$ orbital family extends to a larger radius with a higher energy. The segment located between $x=0$ and $0.3$ inside $\xtwo$ ($\xthree$) orbital family is the orbits generated by the BH. For the detailed shape of these orbits we refer the readers to \citet{com81}, \citet{con89}, \citet{ath92a}, \citet{sell93} and \citet{kim12b}.

\subsection{General evolution}
Our canonical model O33C24B10 is nearly identical to the standard model in \citet{kim12a} except that $\cs$ is increased from $5$ to $8\kms$, and they used CMHOG2 code in cylindrical geometry. The overall evolution of both models is very similar, despite the two codes being quite different, thus demonstrating that the simulation results are robust.

We briefly describe the gas evolution in our canonical model, which can also be seen in Figure~\ref{fig:evo}. At $t=0\Gyr$, gas is distributed uniformly over the whole simulation domain. As the bar grows, gas begins to gather in ridges at the leading side of the bar. When the bar potential becomes strong enough, the ridges develop into sharp dust-lane shocks. These shocks are curved and thus prone to the wiggle instability that occurs due to accumulation of potential vorticity at the distorted shock fronts (\citealt{wad04,kim12a,kim14}). Gas loses its angular momentum when it hits the dust-lane shocks, falling inward to form a nuclear ring at the position where the centrifugal force balances the gravity. Small dense clumps produced at the dust-lane shocks due to the wiggle instability also fall into the ring, making it fairly inhomogeneous (\citealt{kim12a}).

The bisymmetric bar potential induces weak $m = 2$ perturbations in the central regions. The perturbed gas elements follow slightly elliptical orbits, forming spiral structures whose shape depends critically on the sign of $d(\Omega-\kappa/2)/dR$, such that they are leading (trailing) where $d(\Omega-\kappa/2)/dR$ is positive (negative) (\citealt{com96}). Here $\Omega^2\equiv R^{-1}d\Phi_{\rm ext}/dR$ and $\kappa^2\equiv R^{-3}d(R^4\Omega^2)/dR$ denote the angular and epicyclic frequencies (\citealt{bin08}). In our models the nuclear spiral structures appear in the early time, with the shape similar to the previous simulations (e.g. \citealt{eng00,mac04a,mac04b,tha09}). However, these nuclear spirals are destroyed later due to the interaction with nuclear rings. When $t\geq 600\Myr$, the gas flow near the nuclear ring region reaches a quasi-steady state, although the dust-lane shocks and nuclear spirals decay slowly. These results are overall consistent with the previous low-resolution simulations with a point symmetry relative to the galaxy center (e.g., \citealt{ath92b,eng97,reg03}).

Evolution of other models with different $\Omb$ and $\rhobul$ is qualitatively similar except that nuclear rings are absent if $\Omb$ is too large or $\rhobul$ is too small. Figure \ref{fig:snap} displays snapshots of gas surface density in logarithmic scale at $t=800\Myr$ from the third set of models in Table \ref{tbl:model}. Note that a nuclear ring tends to be larger for smaller $\Omb$ and/or larger $\rhobul$. We will explore this trend more quantitatively in the following sections.

\begin{figure}[!t]
\epsscale{1.0} \plotone{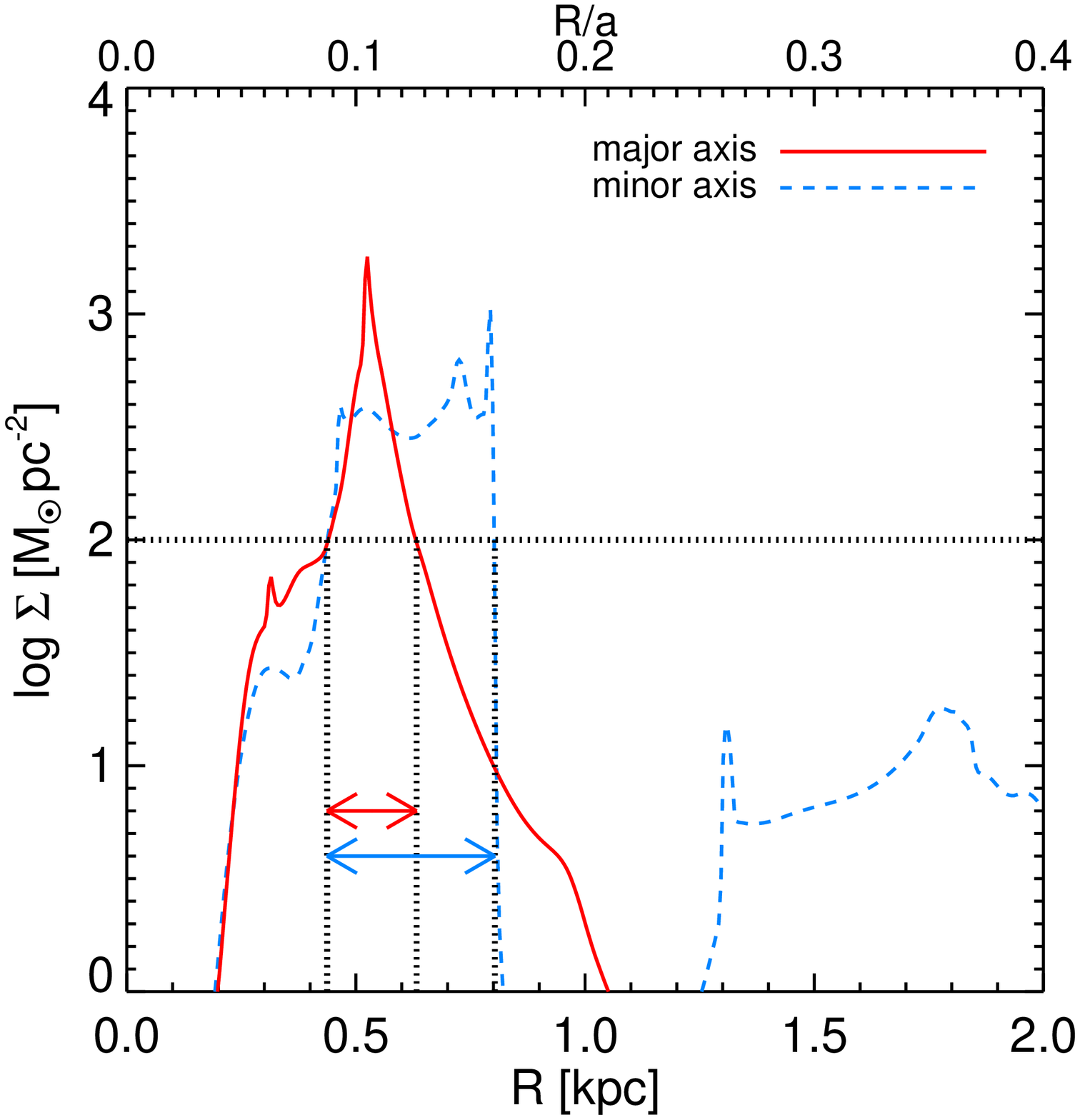}
\caption{Density cut in the canonical model O33C24B10 along bar minor (blue) and major (red) axes at $t=940\Myr$. The dotted line indicates the threshold surface density $\Sigma=100\surf$ for nuclear rings. The short lines with arrows on both ends indicate the width of the ring in the major and minor axes.
\label{fig:ringprofile}}
\vspace{0.2cm}
\end{figure}

\subsection{Quantitative analysis of nuclear rings}

Unlike nuclear spirals and dust lanes that decay over time, a nuclear ring is a robust feature that lasts for a long period of time. We run the canonical model up to $3\Gyr$ and confirm that the nuclear ring still persists, with its shape and size relatively unchanged. This holds true even when nuclear rings are subject to star formation and supernova feedback (\citealt{sk13,sk14}), demonstrating that the lifetime of nuclear rings could be at Gyr timescale.

The nuclear rings in our simulations are roughly elliptical. Figure \ref{fig:ringprofile} plots the density cut profiles along the bar major and minor axes in the canonical model. In this paper we define a nuclear ring exists where the gas surface density exceeds $100\surf$, since the gas surface density in the nuclear ring region of real galaxies is between $100$ to $1000\surf$ (\citealt{kk04}). The surface density in the nuclear ring region of our canonical model can be as high as $1500\surf$. Note that the nuclear ring is denser and narrower along the bar major axis than the minor axis, which is most likely due to the convergence of the outer $\xtwo$ orbits on the major axis (e.g. \citealt{reg03}). The extent and thickness of the ring are shown by the arrows in Figure \ref{fig:ringprofile}. We then examine the relation between the size and shape of the nuclear ring and the galactic parameters discussed in Section~\ref{sec:modelsetup}. Table \ref{tbl:model} lists inner and outer nuclear ring radii along bar major and minor axes in each simulation. Note that model O43C16B10 does not form a nuclear ring. The elliptical structure in the center of this model has a very low surface density, therefore it does not satisfy our nuclear ring definition ($\Sigma\geq100\surf$). Also there is no $\xtwo$ orbital family in this model.

For simplicity, we name the nuclear rings according to their morphology. That is, the rings whose shape is similar to the $\xtwo$ orbital family are named \textit{$\xtwo$-type rings}; these rings are almost round and slightly elongated along the bar minor axis. Several examples of $\xtwo$-type rings can be seen in Figure \ref{fig:snap}, such as models O33C24B10 and O43C40B10. For the nuclear rings highly elongated along the bar major axis which probably follow the $\xone$ orbital family, we name them \textit{$\xone$-type rings} (models O33C16B10 and O43C24B10). The definitions are the same as in \citet{kim12b}.

\begin{figure}[!t]
\epsscale{1.0} \plotone{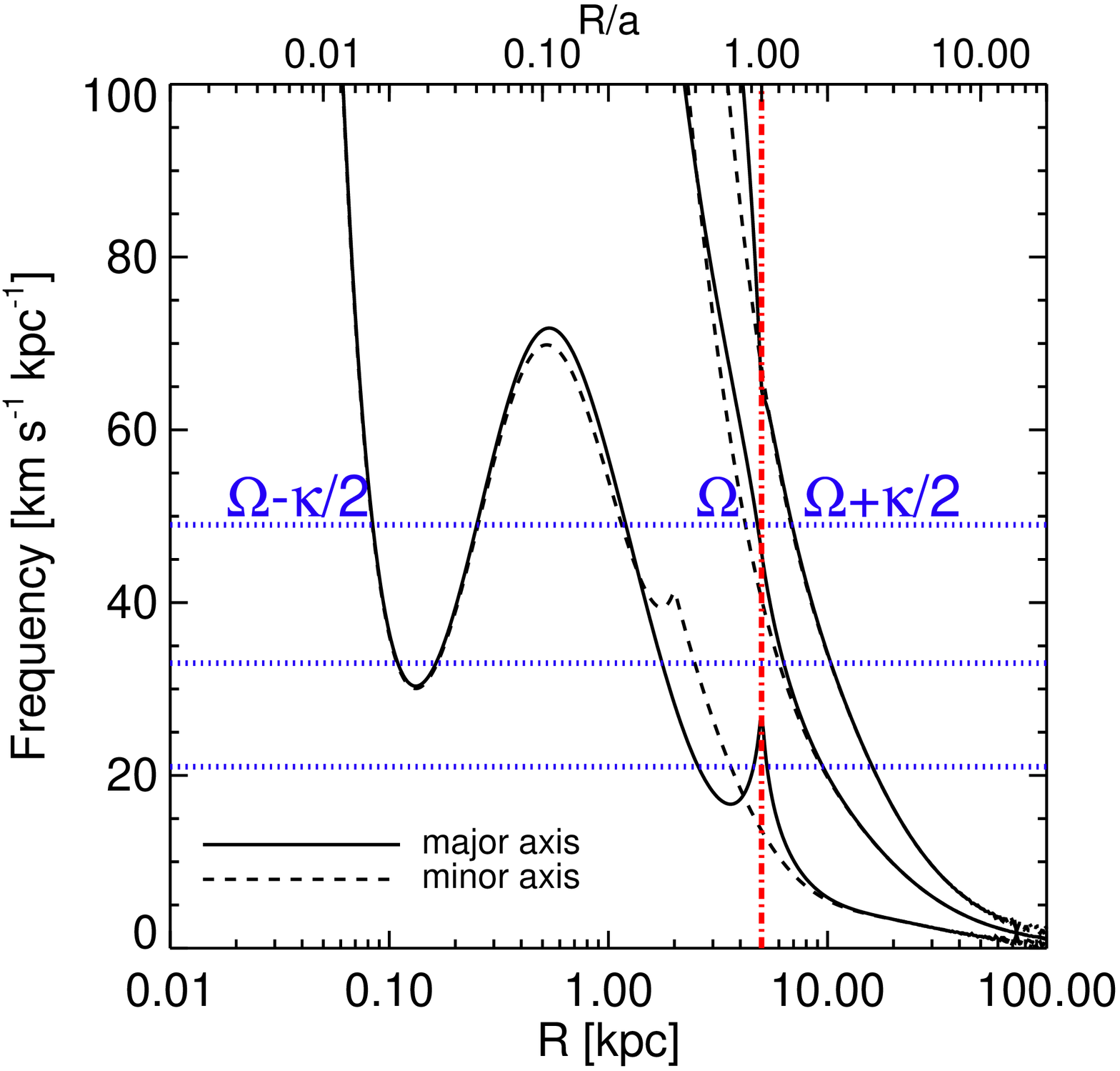}
\caption{Frequency curves for models O21C24B10, O33C24B10, and O49C24B10. The solid and dashed lines represent $\Omega-\kappa/2$ (leftmost curves), $\Omega$ (middle curves) and $\Omega+\kappa/2$ (rightmost curves) along the bar major (solid) and minor (dotted) axes, respectively. The horizontal dotted blue lines correspond to $\Omb= 49, 33 ,21\freq$ from top to bottom. The vertical dot-dashed red line indicates the bar length $a$, and the upper $x$-axis labels $R/a$. Note that the number of ILRs depends on $\Omb$.
\label{fig:omegaresonance}}
\vspace{0.2cm}
\end{figure}

\begin{figure}[!t] 
\epsscale{1.0} \plotone{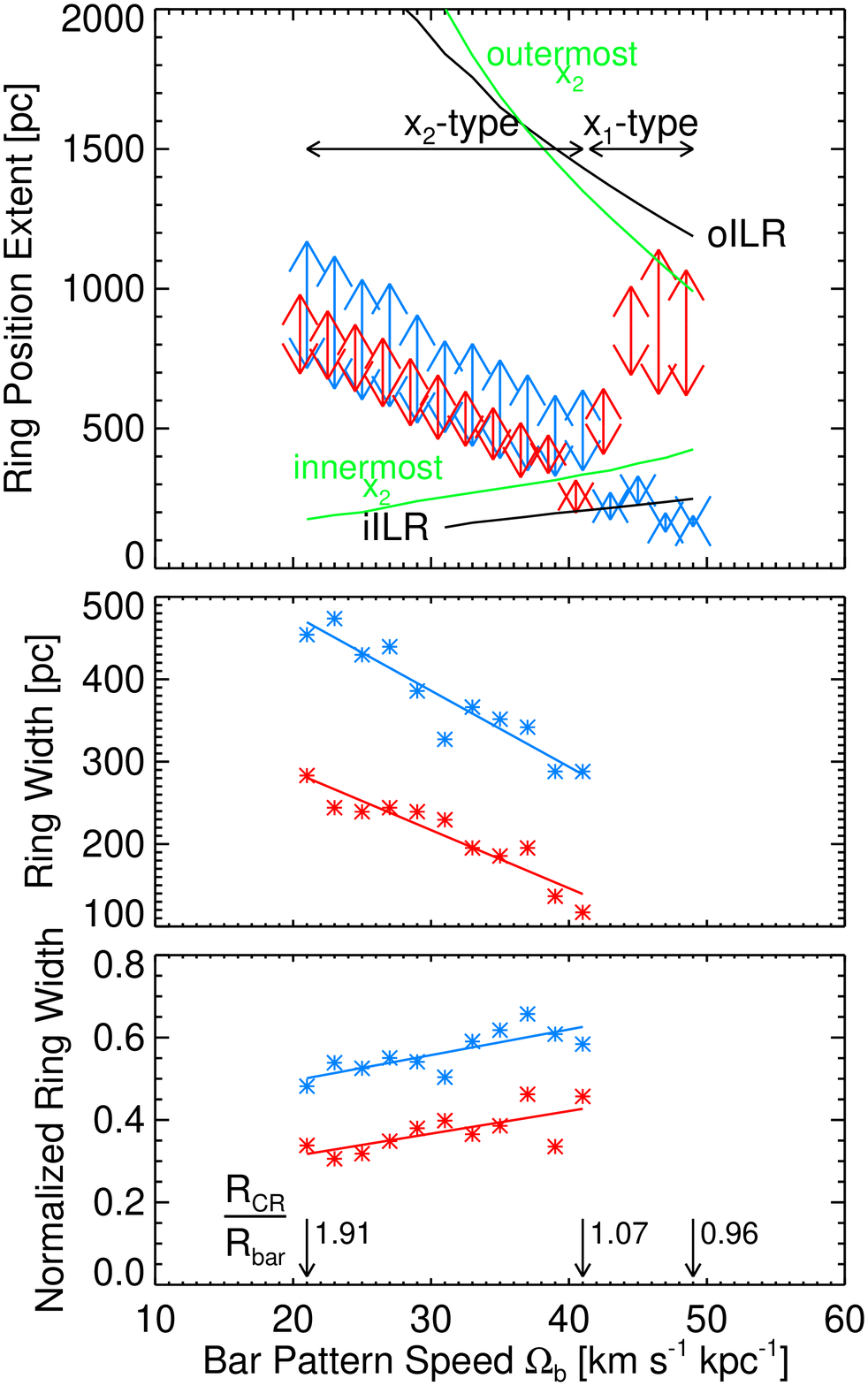}
\caption{Upper panel: relationships between nuclear ring size and thickness with $\Omb$. The arrows indicate the inner and outer radii of the rings along the bar minor (blue) and major (red) axis as defined in Figure \ref{fig:ringprofile}. Galaxies with $\Omb>\Ombc=41\freq$ have $\xone$-type rings, while those with $\Omb<\Ombc$ have $\xtwo$-type rings. The black solid lines mark the positions of the oILR and iILR: galaxies with $\Omb < 31\freq$ have only one ILR. Green lines indicate innermost and outermost $\xtwo $ orbits along the bar minor axis. Middle and lower Panels: absolute and normalized nuclear ring thickness vs. $\Omb$. The asterisks are numerical results and the solid lines are the fits.
\label{fig:ringvsomegab}}
\vspace{0.2cm}
\end{figure}

\subsection{Varying pattern speed}

\subsubsection{The relation between nuclear ring size and $\Omb$}

The pattern speed $\Omb$ is one of the most important bar parameters. We first study the relation between ring properties and $\Omb$ in our models. We vary $\Omb$ from $21$ to $49\freq$, corresponding to the co-rotation resonance radius $\RCR=1.91a$ to $0.96a$\footnote{While $\Omb$ higher than $47\freq$ corresponds to $\RCR/a$ less than unity, which is not self-consistent due to the fact that no $x_1$ stellar orbits (which are the backbones of the bar) can extend beyond the co-rotation radius (\citealt{con80}), these high $\Omb$ models are included to study the effect of $\Omb$ in a systematic way.}. Figure \ref{fig:omegaresonance} plots the characteristic angular frequencies $\Omega-\kappa/2$, $\Omega$, and $\Omega+\kappa/2$ along the bar major and minor axes by solid and dashed lines in these models, respectively. The middle horizontal dotted line represents our canonical model with $\Omb=33\freq$ ($\RCR=1.2a$). Strictly speaking, the resonance radii that we have described are only valid for a weak non-axisymmetric perturbation. For the strong bars, this estimation of the resonance radii by the linear perturbation theory is only indicative. The bar strength parameter for the canonical model is $0.242$ as a reference ($\Qb$ is same for all models in the first set of Table \ref{tbl:model}), which corresponds to a strong bar\footnote{$\Qb$ is defined as the maximal ratio of the tangential force (mainly due to the non-axisymmetric bar potential) to the azimuthally-averaged radial force in $\Phi_{\rm ext}$ (e.g., \citealt{com81,lau04,com10}).}.

We also calculate the $\xtwo$ orbital family in these models. The extent of the $\xtwo$ orbital family may be used to generalize the concept of an ILR to finite amplitude perturbations. However, \citet{van82} found that there are strong bars with no $\xtwo$ orbital family, although ILRs still exist as predicted by the epicyclic theory. Therefore the existence of an ILR is a necessary but insufficient condition for the existence of the $\xtwo$ orbital family.

Figure \ref{fig:omegaresonance} shows that changing $\Omb$ affects both iILR and oILR. For $\Omb$ varing from $21$ to $29\freq$ models have only one ILR if we ignore the artificial cusp introduced by the sharp density truncation at the edge of the Ferrers bar. On the other hand, when $\Omb$ is larger than $31\freq$, models have three ILRs. The distance between iILR and oILR decreases as $\Omb$ increases.

Figure \ref{fig:ringvsomegab} gives the size and thickness of the rings as a function of $\Omb$ in our first set of simulations. In the upper panel, the black solid lines draw the radii of the oILR and iILR, while the green lines mark the radii of the outermost and innermost $\xtwo$ orbits along the bar minor axis. Note that the shape of nuclear rings depends on $\Omb$. For $21\freq<\Omb<41\freq$ ($1.07a<\RCR<1.91a$), the nuclear rings are almost circular but slightly elongated along the bar minor axis, so they are $\xtwo$-type rings. The size and thickness of these nuclear rings along the bar minor (blue arrows) and major axis (red arrows) follow nearly a linear relation with $\Omb$, such that a higher $\Omb$ gives a narrower and smaller ring. For these nuclear rings, the upper ends of the blue arrows are higher than those of the red arrows, with an ellipticity\footnote{The ring ellipticity is defined as $\ering=1-b_r/a_r$, where $a_r$ and $b_r$ are the semi-major and semi-minor axes of the ring.} of $\ering\sim0.1$--$0.3$ at the outer boundary. The lower ends of the blue and red arrows are very close to each other, indicating the inner boundary of these rings is nearly circular. The observed ring ellipticities are less than $0.4$ as reported by \citet{com10} and \citet{maz11}, which means our numerical results are consistent with observations. The variation of nuclear ring thickness also shows a trend that the absolute ring thickness decreases with increasing $\Omb$, but the trend is reversed with respect to the normalized ring thickness (the ratio of ring thickness to ring radius). We further explain and discuss these trends in Section~\ref{sec:explaination}.

\begin{figure}[!t]
\epsscale{1.2} \plotone{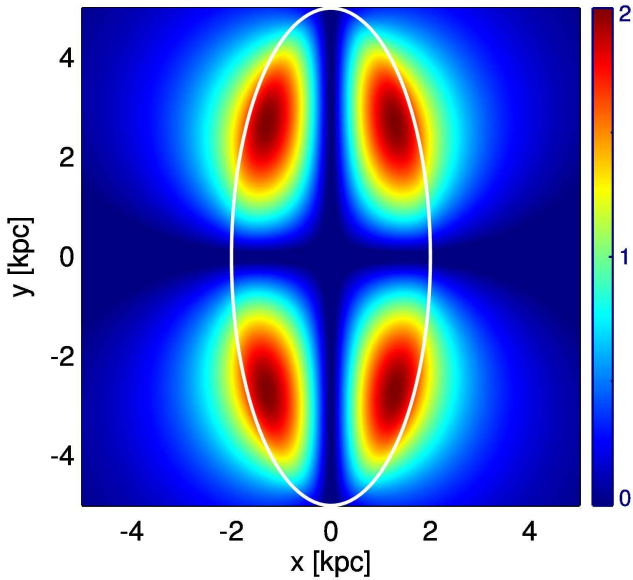}
\caption{$\lvert \mathbf{r} \times \mathbf{F} \rvert$ for our Ferrers bar model. The white ellipse denotes the bar outline. We see that the region where the largest $\lvert \mathbf{r} \times \mathbf{F} \rvert$ occurs is mainly at the inner edge of the bar.
\label{fig:bartorque}}
\vspace{0.2cm}
\end{figure}

\begin{figure*}[!t]
\epsscale{1.1} \plotone{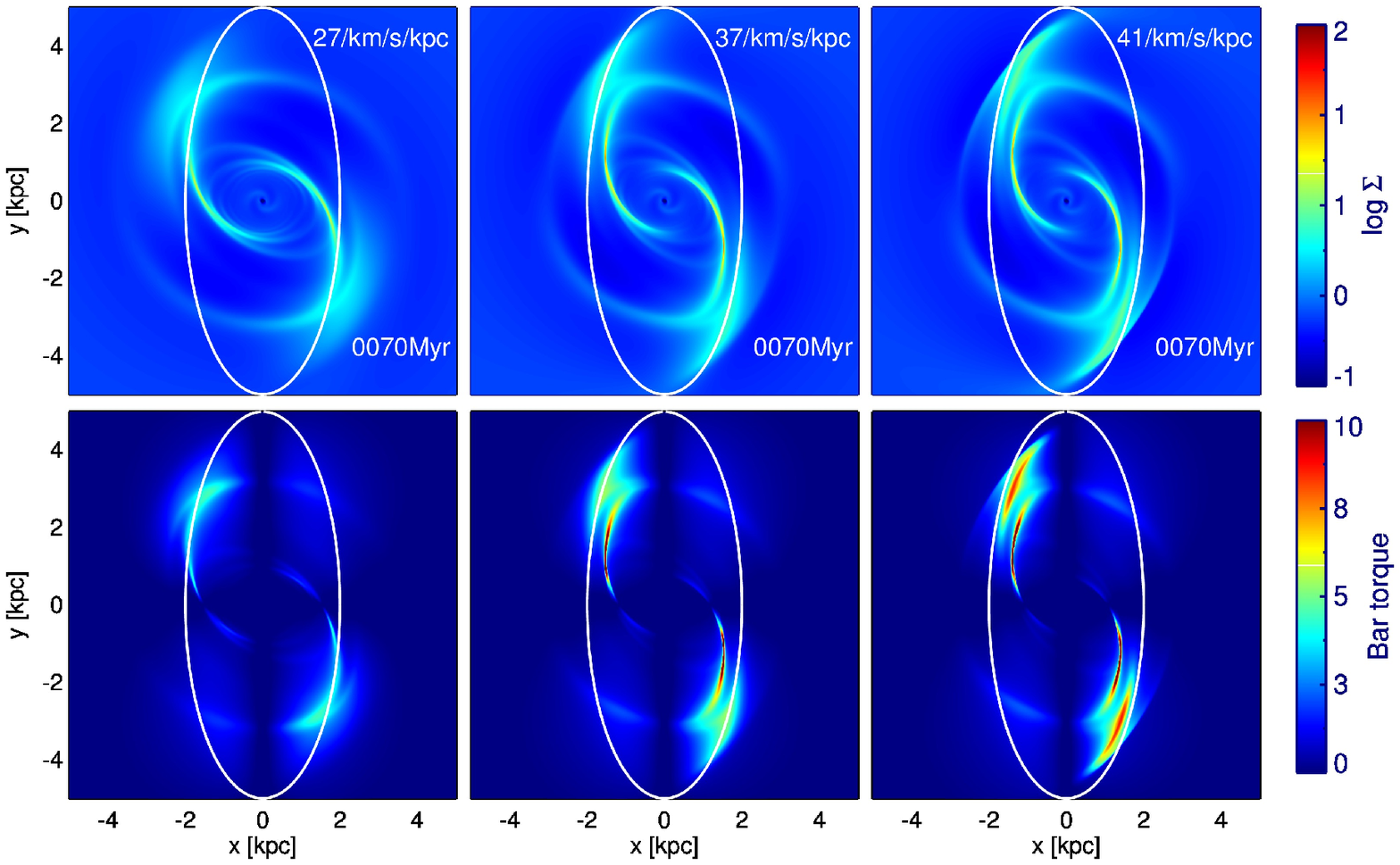}
\caption{Gas surface density $\Sigma$ (upper panels) and the corresponding bar torque $\Sigma \lvert \mathbf{r} \times \mathbf{F} \rvert$ on the gas (lower panels) for 3 different $\Omb$ models ($27, 37, 41\freq$). The white ellipse denotes the bar outline. We see that dust-lane shocks (overdense ridges) in models with high $\Omb$ tend to be located in the regions where the non-axisymmetric bar torque is large. This means that gas in higher $\Omb$ models experiences a larger bar torque compared to those in lower $\Omb$ models, leading to higher angular momentum loss in gas.
\label{fig:torquecompare}}
\vspace{0.2cm}
\end{figure*}

When $\Omb\geq41\freq$, the rings suddenly become very elongated along the bar major axis, which can be seen by the large separation between the red and blue arrows (thus these are $\xone$-type rings). However, $\xone$-type rings are almost never seen in observations (e.g., \citealt{com10}). According to this figure, we conclude that $\xtwo$-type rings can only exist in our models when $\Omb \leq \Ombc=41\freq$ (corresponding to $\RCR=1.07a$). If $\Omb$ is larger than $\Ombc$, then the rings are very elliptical along bar major axis and the linear correlation between $\Omb$ and nuclear ring size breaks down. For the models with $\Omb \leq \Ombc$, all the nuclear rings are $\xtwo$-type and their sizes correlate well with $\Omb$, although these models have different number of ILRs (3 ILRs for $\Omb \geq 31\freq$ and 1 ILR for $\Omb<31\freq$, see Figure \ref{fig:omegaresonance}). Therefore the correlation between $\Omb$ and $\xtwo$-type nuclear ring size is not affected by the number of the ILRs, which may imply that reading ILR directly off the frequency curves is not accurate and does not give accurate information on the ring position, as we cautioned earlier. We remark that $\Ombc=41\freq$ found here is valid only for our current set of model parameters: the critical patten speed depends on the bar and galaxy parameters (for more discussions see Section \ref{sec:varyingdensity} and \ref{sec:discussion}).

Figure \ref{fig:ringvsomegab} also shows that the allowed radial range for the $\xtwo$ orbits follows nicely with the location of the ILRs. The formation of an $\xone$-type ring is due to the limit set by the innermost $\xtwo$ orbit. If the bar drives gas flow inward to a radius that is smaller than the innermost $\xtwo$ orbit, then there are no stable $\xtwo$ orbits to support the inflow gas. Therefore the gas would instead follow the $\xone$ orbital family to remain quasi-steady, since its orbital energy is higher than that of the $\xtwo$ orbital family for a given angular momentum and it extends all the way to the center (see Figure \ref{fig:orbitfamily}). We see that in Figure \ref{fig:snap} and \ref{fig:ringvsomegab}, all $\xone$-type rings are well inside the innermost $\xtwo$ orbit along the bar minor axis. This may explain why there is a break for the trend of nuclear ring sizes on Figure \ref{fig:ringvsomegab}. We will elaborate further on this point in Section~\ref{sec:discussion}.

\subsubsection{The torque exerted on the gas by the bar}

We try to explain why higher $\Omb$ models tend to have smaller rings. The non-axisymmetric bar potential produces shocks in the gas flows that would otherwise follow galaxy rotation. Owing to a larger velocity relative to the bar potential, shocks are stronger in models with relatively smaller $\Omb$. Stronger shocks due to the bar potential tend to form farther downstream from the potential minimum, i.e., the bar major axis (\citealt{hq11}), analogous to the cases of spiral shocks (\citealt{kim02,kim06,git04,kk14}). In Figure \ref{fig:bartorque} we plot the absolute value of $(\mathbf{r} \times \mathbf{F})$ for our canonical Ferrers bar model, where $\mathbf{F}$ is the gravitational force due to the bar. The white ellipse represents the bar outline. We see that the maxima of $\lvert \mathbf{r} \times \mathbf{F} \rvert$ occur near the inner edge of the bar. In Figure \ref{fig:torquecompare} we plot the gas surface density $\Sigma$ and the corresponding bar torque $\Sigma \lvert \mathbf{r} \times \mathbf{F} \rvert$ on gas at $t=80\Myr$ for models with $3$ different $\Omb$ ($27, 37, 41\freq$)\footnote{In these models the growth time of the bar is fixed at $100\Myr$ in order to facilitate the comparison.}. The model with a higher $\Omb$ produces overdense ridges with a smaller offset from the bar major axis. In Figure \ref{fig:torquecompare} we see that dust-lane shocks in models with high $\Omb$ tend to be located in the regions where the non-axisymmetric bar torque is larger, while those in models with low $\Omb$ are located in the regions with smaller bar torque. This result means that the gas in higher $\Omb$ models experiences a stronger bar torque.

Lower $\Omb$ models form larger rings because the gas loses less angular momentum due both to shocks and to the bar torque. On the other hand, models with a relatively lower $\Omb$ have stronger shocks, but these models still apparently lose less angular momentum in total (also see discussions in Section~\ref{sec:explaination}). This suggests that the bar torque may play a more important role in removing angular momentum than the shocks. We will study the process of the angular momentum removal in greater depth in the future.

\begin{figure}[!t]
\epsscale{1.0} \plotone{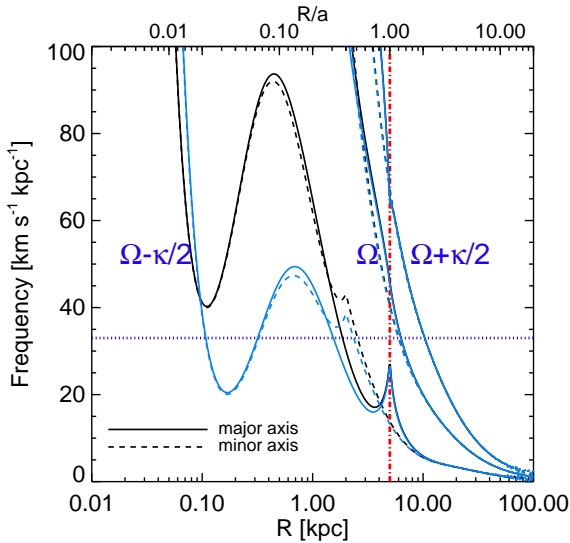}
\caption{Frequency curves for models O33C12B10 (blue) and O33C40B10 (black). Similar to Figure \ref{fig:omegaresonance}, the solid and dashed lines represent $\Omega-\kappa/2$ (leftmost curves), $\Omega$ (middle curves), and $\Omega+\kappa/2$ (rightmost curves) along bar major and minor axes. When we change $\rhobul$, we also modify the scale length of bulge to keep the total mass within $10\kpc$ constant, so that different $\rhobul$ only affects the central region ($R \simlt 2\kpc$). The horizontal dotted blue line corresponds to $\Omb= 33\freq$ for these models. Note that $\rhobul$ changes the number of ILRs.
\label{fig:centralresonance}}
\vspace{0.2cm}
\end{figure}

\begin{figure}[!t] 
\epsscale{1.0} \plotone{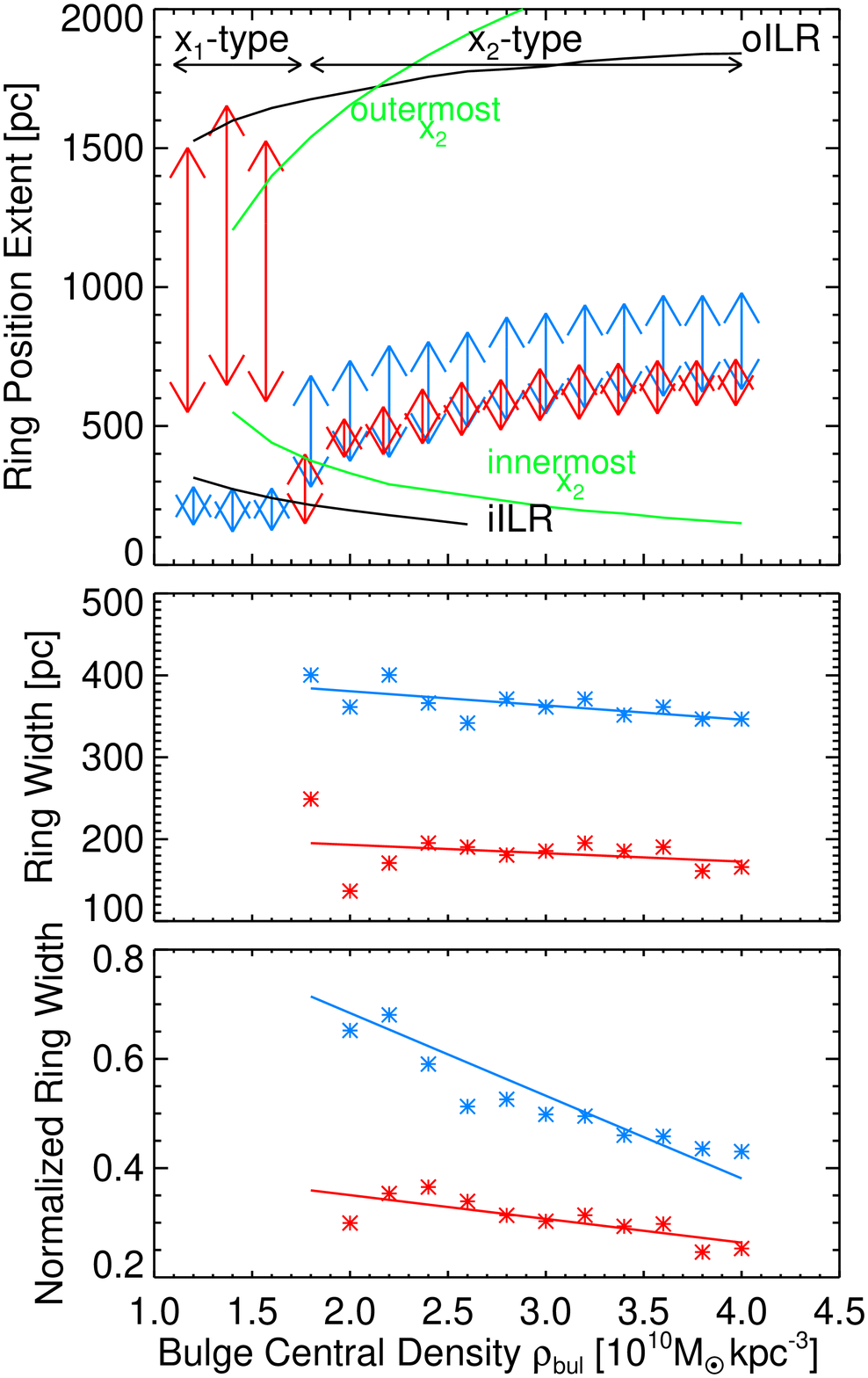}
\caption{Upper panel: relationships between nuclear ring size and thickness with different $\rhobul$. The arrows indicate the inner and outer radii of the rings along the bar minor (blue) and major (red) axis as defined in Figure \ref{fig:ringprofile}. Models with $\rhobul > \rhobulc = 1.8 \times 10^{10}\dunit$ tend to have an $\xtwo$-type ring, while those with less central density tend to have an $\xone$-type ring. The black solid lines mark the locations of the oILR and iILR, the second of which ceases to exist when $\rhobul > 2.6 \times 10^{10}\dunit$. Green lines indicate the innermost and outermost $\xtwo$ orbits along the bar minor axis. Note that the leftmost model in the upper panel (model O33C12B10) does not have the $\xtwo$ orbital family. Middle and lower Panels: absolute and normalized nuclear ring thickness vs. $\rhobul$. The asterisks are the numerical results and the solid lines are the fits.
\label{fig:ringvscentralden}}
\vspace{0.2cm}
\end{figure}

\subsection{Varying central density}
\label{sec:varyingdensity}

The bulge component defines the Hubble sequence of the host disk galaxy, and it plays an important role in secular evolution. N-body simulations further suggest that the bulge and bar can also interact with each other (e.g., \citealt{ath02,ss04}). Therefore, the influence of bulge mass or bulge type/profile on the gas flow in barred galaxies should also be considered. In this study we experiment with the impact of the bulge central density $\rhobul$ on nuclear rings. We vary $\rhobul$ from $1.2 \times 10^{10}$ to $4.0 \times 10^{10}\dunit$. In order to keep the total mass of the bulge within $10\kpc$ constant ($\Mbul$ within $10\kpc$ is fixed to $3.37\times10^{10}\Msun$, which is about $36\%$ of the total bulge+bar mass within $10\kpc$), the scale length of bulge $R_b$ is changed accordingly from $0.467\kpc$ to $0.306\kpc$, with smaller $R_b$ corresponding to a more concentrated bulge. The variation of $\rhobul$ and $R_b$ barely affects the bar strength, giving $\Qb=0.246$ for model O33C12B10 and $\Qb=0.234$ for model O33C40B10.

Figure \ref{fig:centralresonance} shows the frequency curves in these models. Different $\rhobul$ generates a different number of resonances, similar to the case of varying $\Omb$. In Figure \ref{fig:ringvscentralden} we show the variations of the ring size and thickness with $\rhobul$. Again, the black solid lines indicate the location of the oILR and iILR. The ring type depends not only on $\Omb$ but also on $\rhobul$. We find that lower $\rhobul$ prefers $\xone$-type rings and higher $\rhobul$ tends to form larger $\xtwo$-type rings with a critical bulge central density of $\rhobulc=1.8 \times 10^{10}\dunit$. This result suggests that we need a relatively massive and concentrated bulge to support the formation of $\xtwo$-type rings, which is consistent with previous simulation results (e.g., \citealt{ath92b,reg03}). According to the critical model (O33C18B10), if the enclosed bulge mass at $1\kpc$ is less than $\sim15\%$ of the total enclosed mass at $5\kpc$, then it is hard to form an $\xtwo$-type ring. The absolute thickness of nuclear rings is not sensitive to $\rhobul$, both along the bar major and minor axis. In fact the absolute thickness exhibits a slightly negative correlation with $\rhobul$, as shown in the middle panel of Figure \ref{fig:ringvscentralden}. The normalized ring thickness shows a slightly stronger negative correction with $\rhobul$. These correlations may be examined with future observations.

The correlation of $\rhobul$ with the nuclear ring size becomes flat on the high $\rhobul$ end, which may imply that $\rhobul$ higher than $4.0 \times 10^{10}\dunit$ generates nuclear rings similar to that in model O33C40B10. We conclude that $\xtwo$-type rings can only exist in our models when $\rhobul > \rhobulc=1.8 \times 10^{10}\dunit$. Of course, $\rhobulc$ may be different if we use other types of bulge profiles. Considering the effects of the $\Omb$, this critical value can be smaller with a smaller $\Omb$ than $\Ombc=41\freq$ as we do see an $\xtwo$-type ring in model O25C16B10 (Figure \ref{fig:snap}), and the critical value of $\Omb$ can be higher with higher $\rhobul$ than $\rhobulc$. Again we see that the correlation between $\rhobul$ and $\xtwo$-type ring size is not affected by the number of iILRs either, which may be also due to the inaccurate estimation of resonances for a strong bar.

The innermost and outermost $\xtwo$ orbits are plotted as green lines in Figure \ref{fig:ringvscentralden}. The break between the $\xone$- and $\xtwo$-type rings here is still due to the boundary set by the innermost $\xtwo$ orbit. If we decrease $\rhobul$ below $\rhobulc$, the $\xtwo$ orbits can no longer exist as the center is not dense enough. The radius of the innermost $\xtwo$ orbit increases with decreasing $\rhobul$, and an $\xone$-type ring may form when the gas falls inside the innermost $\xtwo$ orbit.

\begin{figure*} 
\epsscale{1.1} \plotone{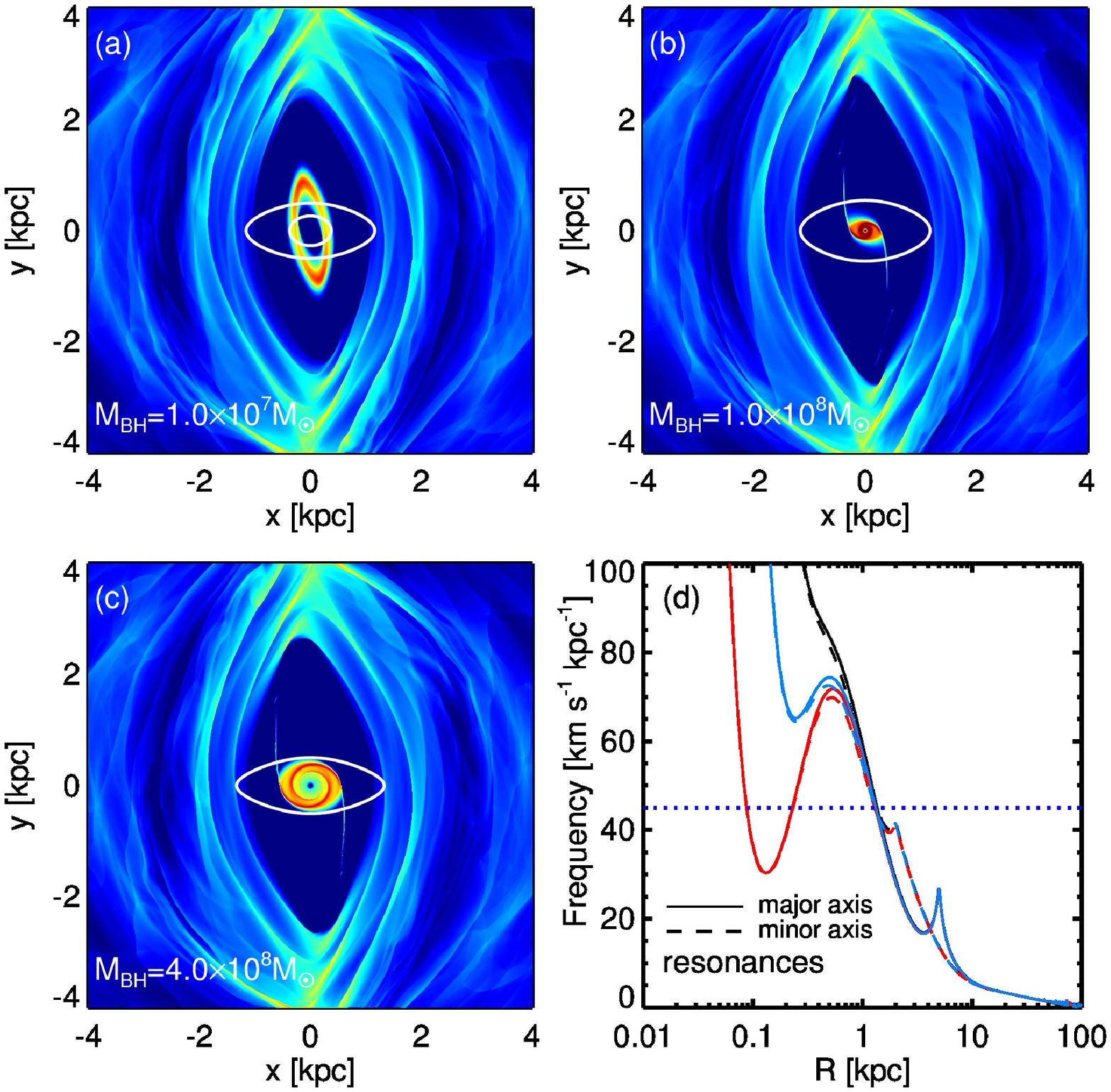}
\caption{Gas flow patterns for (a) O45C24B10 with $\MBH=1 \times 10^7\Msun$, (b) O45C24B100 with $\MBH=1 \times 10^8\Msun$, (c) O45C24B400 with $\MBH=4 \times 10^8\Msun$ at $t = 950\Myr$, and (d) the corresponding frequency curves. In (a), the white ovals plot the innermost and outermost $\xtwo$ orbits.  In (b) and (c), only the outermost $\xtwo$ orbit is given, since the $\xtwo$ orbits extend to the center due to a massive ``BH''. In (d), the red, blue and black curves give $\kappa-\Omega/2$ for models with O45C24B10, O45C24B100, and O45C24B400, respectively. The horizontal dotted blue line marks $\Omb = 45\freq$ for these models. Note that it is very hard to identify whether the nuclear structure in panel (b) and (c) is an $\xtwo$-type ring or a pair of tightly-wound nuclear spirals.
\label{fig:BHeffect}}
\vspace{0.2cm}
\end{figure*}

\subsection{Effects of a massive central ``black hole"}

\citet{kim12a} has demonstrated that adding a massive BH does not greatly affect the shape and size of $\xtwo$-type rings, so we investigate the influence of the BH in those galaxies with $\xone$-type rings. We use models O45C24B10, O45C24B100 and O45C24B400 together to demonstrate our results. Figure \ref{fig:BHeffect} shows that model O45C24B10 forms an $\xone$-type ring with outer semi-major and semi-minor axes of $~600\pc$ and $150\pc$. The gas forms an $\xtwo$-type ring with an averaged radius of $\sim200\pc$ in model O45C24B100 and of $\sim500\pc$ in model O45C24B400. The corresponding frequency curves and density distributions of these three models are shown in Figure \ref{fig:BHeffect}. We emphasize that the BHs here are unrealistically massive; in real galaxies the mass of the BH is about $\sim0.15-0.5\%$ of the bulge mass $\rm M_{\rm bul}$ according to the $\rm M_{\rm BH}-\rm M_{\rm bul}$ relation (e.g., \citealt{mag98,gul09,kh13}). The bulge mass within $2\kpc$ here is $4.65\times10^{9}\Msun$ so that the BH should have a mass similar to $~1-3\times10^{7}\Msun$. Since we use a much more massive BH than $0.5\%\rm M_{\rm bul}$, the ``BH'' actually acts as another ultra-compact bulge in practice.

Still, our ``BH'' potential dominates only the very inner part in the galaxy, as revealed in Figure~\ref{fig:rotcurve} and \ref{fig:BHeffect}(d). The main difference among these models with different $\MBH$ is the presence/absence of an iILR (or innermost $\xtwo$ orbit). By removing the iILR, models O45C24B100 and O45C24B400 produce $\xtwo$-type rings, indicating that the formation of $\xone$-type rings may be related to the presence of an iILR or the existence of an innermost $\xtwo$ orbit, consistent with previous results. However, varying the mass of the ``BH'' barely changes the location of oILR, as shown in Figure \ref{fig:BHeffect}(d). This result shows that one can generate quite different nuclear rings while keeping oILR unchanged, which is a counter-example to the simulations in \citet{she00} who found that the radius of the nuclear ring was proportional to the radius of the oILR. Therefore at least in some situations the size of nuclear rings does not rely only on the location of oILR. Combining with the results of the last subsection, we conclude that the locations of $\xtwo$-type rings may not be well predicted by the ILRs directly read off the frequency curves, which is similar to the findings in \citet{van09} and \citet{pin14}. Figure \ref{fig:BHeffect} shows that a central ``BH'' with proper mass could help to form ultra-compact nuclear rings ($\Rring/a\leq0.1$).

We have shown previously that in order to form an $\xtwo$-type ring in the central part, we need $\xtwo$ orbits to extend all the way to the center (in other words, we need the radius of the innermost $\xtwo$ orbit to be as small as possible). This can be achieved by a massive ``BH'' potential. As we see in Figure \ref{fig:BHeffect}, a massive ``BH'' removes the inner boundary of the $\xtwo$ orbital family. In such a galactic potential we could form tiny $\xtwo$-type rings without worrying about the formation of $\xone$-type rings.

\section{Discussion}
\label{sec:discussion}

\begin{figure*}
\epsscale{1.1} \plotone{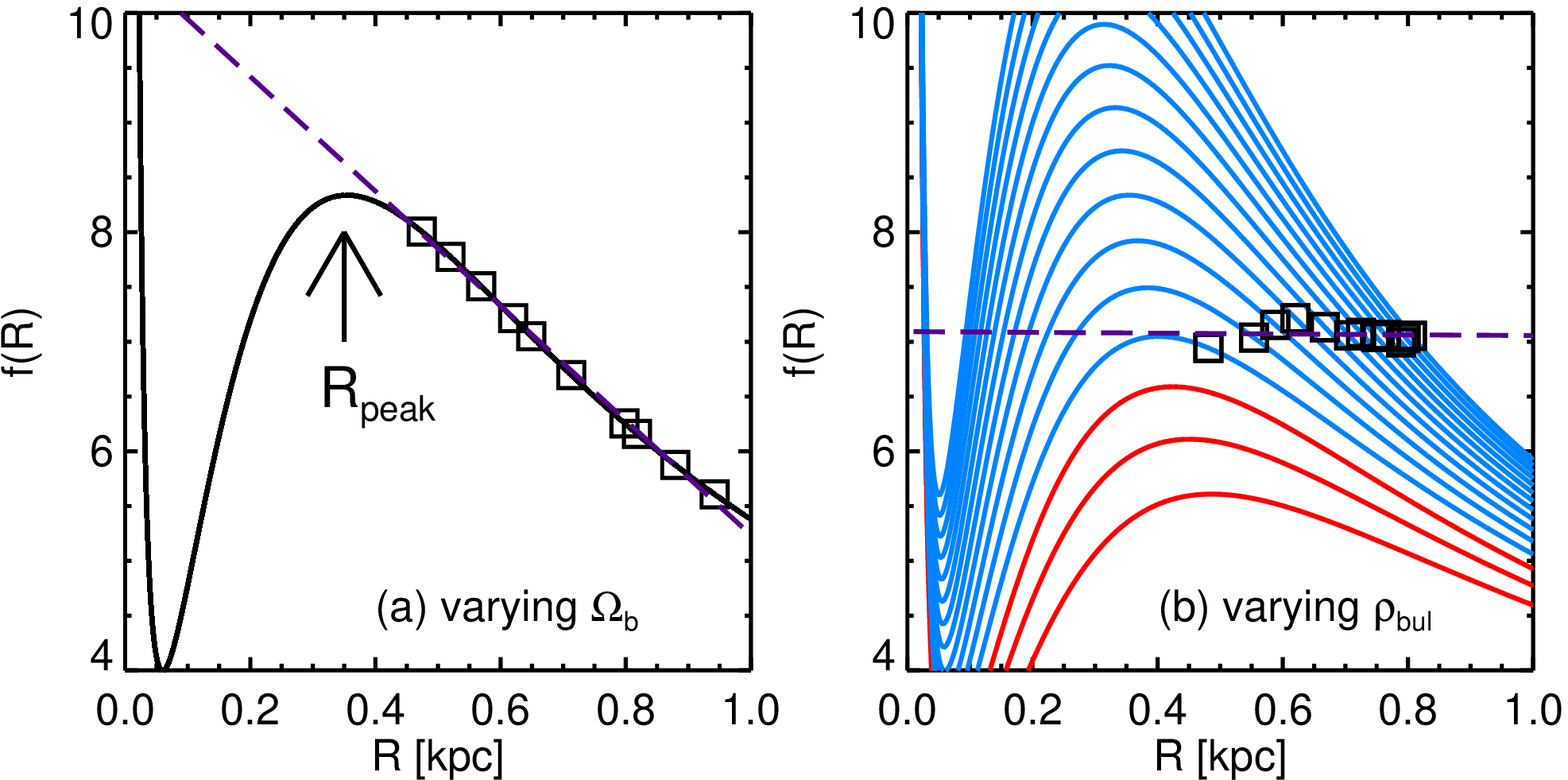}
\caption{Relation between nuclear ring radius and the normalized centrifugal force profile $f(R)$. The black line in panel (a) represents the $f(R)$ curve of the models with varying $\Omb$, while the red and blue $f(R)$ curves in panel (b) represent the models that have $\xone$- and $\xtwo$-type rings with varying $\rhobul$. The open black squares give the numerical results for the $\xtwo$-type rings, while the long-dashed purple lines are their fits. The black arrow in panel (a) indicates the position of $R_{\rm peak}$. $R_{\rm peak}$ is generally smaller than $R_{\rm to}$.
\label{fig:ring_prediction}}
\vspace{0.2cm}
\end{figure*}

\subsection{The formation mechanism of nuclear rings}
\label{sec:fomationmechanism}

The explanation for the formation of a nuclear ring is still under debate. It has often been argued that the location of a nuclear ring should be linked to the ILRs, since between the co-rotation radius and ILRs the potential drives mass inward (\citealt{com96}). However, at exactly which radius (relative to the ILRs) a nuclear ring forms is still not well understood. As Equation \eqref{eq:mom} suggests, the gas flow is not only determined by the external gravitational field $\Phi_{\rm ext}$ and the pattern speed $\Omb$, which govern all the resonances, but also affected by the sound speed $\cs$. \citet{kim12a} have shown that a larger $\cs$ tends to yield a smaller nuclear ring, as larger thermal pressure bends the gas streamlines more strongly, inducing stronger dust-lane shocks and causing a larger amount of angular momentum loss. This implies a more fundamental interpretation, which does not rely only on resonances, is needed to explain the formation of nuclear rings.

We propose that there are three conditions to form a roundish $\xtwo$-type ring. First, the galaxy needs to have enough gas that can lose angular momentum in the center. This may explain why gaseous nuclear rings are rare in early type disk galaxies, as they may have run out of gas (e.g., \citealt{kna05}). Second, the $\xtwo$ orbital family should exist in the galaxy, as it is the backbone of the nuclear ring. Third, exactly where the ring settles is determined by the interplay of the residual angular momentum of the gas after interacting with the potential which is governed by gas dynamics, and the radial distribution of centrifugal force (or $\xtwo$ orbits) which is determined by the mass distribution of the galaxy. We explain the third point in more detail below.

To estimate the total angular momentum removal of gas, we simplify the problem with some assumptions. We assume that the gas around the bar end $R_{\rm bar}$\footnote{In our models, $R_{\rm bar}=a$ and $R_{\rm ring}$ is the ring radius along bar minor axis.} is nearly unperturbed at the beginning and on nearly circular orbits. After the impact of shocks and the bar torque, gas loses angular momentum and flows inward, producing a ring-like structure at $R_{\rm ring}$. The gas in the nuclear ring also follows nearly circular motions. \citet{kim12a,kim12b} showed that a nuclear ring forms where the centrifugal force of the inflowing gas balances the gravity for models with fixed $\Omb$ and $\rhobul$. To see if this is also the case for models varying $\Omb$ and $\rhobul$, we define the normalized centrifugal force relative to the bar end at $R=R_{\rm bar}$ as
\begin{equation}\label{eq:frdefinition}
f(R) \equiv F(R)/F_{\rm bar} = \frac{v(R)^2/R}{v_{\rm bar}^2/R_{\rm bar}}.
\end{equation}
Here $F(R)$ is the centrifugal force at radius $R$, $F_{\rm bar}$ is the centrifugal force at $R=R_{\rm bar}$, and $v_{\rm bar}$ is the circular velocity at $R=R_{\rm bar}$ which is read off the rotation curve. Suppose that the rotation curve is nearly flat with a constant velocity $V_0$\footnote{The assumption of flat rotation curve is indeed simplified, but not critical to our argument. Actually most $\xtwo$-type roundish rings settle on the rising part of the rotation curve. Without loss of generality, we can assume that the velocity at $R$ is $v(R) \propto R^{\alpha}$, thus allowing one to easily show that $F(R) \propto R^{2\alpha-1} \propto (L_{0}-\Delta L)^{(2\alpha-1)/(\alpha+1)}$. A rotation curve generally rises from nearly solid-body rotation, then declines smoothly to the flat(ter) outer part. If $\alpha < 0.5$, $F(R)$ is a monotonically decreasing function of $R$, and $\Delta L$ is still positively correlated with $F$ as shown in Figure~\ref{fig:ring_prediction} and the discussions therein.}, then the centrifugal force at the nuclear ring radius is $F(R_{\rm ring}) = V_0^2/R_{\rm ring} \propto R_{\rm ring}^{-1}$, while the angular momentum loss is $\Delta L \approx V_0(R_{\rm bar}-R_{\rm ring}) = L_0-V_0R_{\rm ring}$, where $L_0 = V_0R_{\rm bar}$, then $F(R_{\rm ring}) \propto (L_0-\Delta L)^{-1}$. The more angular momentum loss $\Delta L$ there is, the larger $F$ is; therefore, $F$ may serve as a rough proxy of $\Delta L$. The total angular momentum removal during the evolution is reflected by the normalized centrifugal force on the nuclear ring $\mring \equiv f(R_{\rm ring}) = F_{\rm ring}/F_{\rm bar}$, where $F_{\rm ring} = F(R_{\rm ring}) = v(R_{\rm ring})^2/R_{\rm ring}$. $F_{\rm ring}$ correlates with the amount of angular momentum loss $\Delta L$ due to shocks and the bar torque, and $F_{\rm bar}$ is a normalizing constant.

To study the distribution of the centrifugal force, we plot the profile of $f(R)$ along bar minor axis in Figure \ref{fig:ring_prediction}. The black squares in panel (a) and (b) are the $\xtwo$-type ring radii along bar minor axis from Figure \ref{fig:ringvsomegab} and \ref{fig:ringvscentralden}. As we assume that gas follows circular motions inside the nuclear ring, those squares should lie on the curves plotted depending on how small the rings are. The dashed lines fit (and extrapolate) the squares.

We denote $R_{\rm peak}$ as the radius where $f(R)$ is peaked, as indicated by the black arrow in Figure \ref{fig:ring_prediction}(a). Since $f(R) \propto R^{2\alpha-1}$ for $v(R) \propto R^{\alpha}$, $R_{\rm peak}$ separates the steeply rising inner part of the rotation curve ($\alpha > 0.5$) and the more slowly rising part in transition to the flat(ter) outer region ($\alpha < 0.5$), but it is not identical to the turnover radius of the rotation curve, $R_{\rm to}$ (usually loosely defined as where $\alpha$ becomes $0$).

Note that all the roundish rings are positioned to the right side of $R_{\rm peak}$, i.e. at the decreasing part of the $f(R)$ curves, indicating that a smaller $x_2$-type ring loses more angular momentum and has a larger $f_{\rm ring}$. We also see that the smallest $\xtwo$-type ring is close to $R_{\rm peak}$. Since the potential does not provide enough centrifugal force for the gas in circular motions at $R<R_{\rm peak}$, one cannot generate $x_2$-type rings smaller than our critical model (i.e., O41C24B10 and O33C18B10) simply by increasing $\Omb$ or decreasing $\rhobul$. This is more evident in Figure \ref{fig:ring_prediction}(b) where the blue and red curves represent the models with $x_2$- and $x_1$-type rings, respectively. An $\xtwo$-type ring cannot form when the dashed line does not intersect with the curve of $f(R)$ before $R_{\rm peak}$. When a more massive ``BH'' is added, on the other hand, it raises the inner part of $f(R)$ and provides a sufficient centrifugal force barrier for an $\xtwo$-type ring at radii that would otherwise form an $x_1$-type ring (see Figure \ref{fig:BHeffect}). If the centrifugal force is insufficient to support the circular motion, then the gas changes its orbits to eccentric ones as it still has non-zero angular momentum (therefore an $\xone$-type ring). This is an alternative explanation for Figure 11 in \citet{kim12b} where they argue that $\xone$-type rings formed due to the velocity of gas are too high to settle on the $\xtwo$ orbital family, and is also an alternative explanation for the $\xtwo$ orbit extent theory we discussed in Section~\ref{sec:simulationresults}. This explanation is simpler as we do not need to compute orbits.

\subsection{Qualitative explanation for the trends of nuclear ring size and thickness}
\label{sec:explaination}

Here we attempt a qualitative physical explanation for the trends of nuclear ring size and thickness (Figure \ref{fig:ringvsomegab} and \ref{fig:ringvscentralden}). The reason that low $\Omb$ simulations form larger rings is because gas loses less angular momentum compared with high $\Omb$ runs, as the bar torque is weaker in low $\Omb$ simulations than the high $\Omb$ ones. For low $\Omb$ simulations, the bar growth time is larger and the gas response time is longer for slow bars; the gas crossing the shock region at later times loses more angular momentum than gas at earlier times as the shocks and bar torque become stronger with time. Thus the nuclear ring is more spatially extended in simulations with a low $\Omb$ simulation than high $\Omb$.

High $\rhobul$ simulations also have sightly smaller gas response time since they have more mass in the central part, so higher $\rhobul$ simulations result in slightly narrower rings. Since the bar strength of these simulations is nearly the same, i.e. $\mring$ is almost constant, the purple dashed line is almost horizontal as is seen in panel (b) of Figure \ref{fig:ring_prediction}. High $\rhobul$ simulations could form larger nuclear rings at a larger radius since they provide the ``required'' centrifugal force barrier first, as gas flows inward from outer part to the center and intersects first with the curves provided by higher $\rhobul$. We emphasize that the angular momentum loss in different $\rhobul$ simulations is nearly the same, making $\mring$ nearly constant for this set of simulations in which the bulge is only an axisymmetric component.

\subsection{Relations between ILRs and $\xtwo$ orbits}

As \citet{van82} suggested, the existence and extent of the $\xtwo$ orbital family can be used to generalize the concept of an ILR for strong bars. We do see that the trends of the $\xtwo$ orbital family and ILRs follow each other closely in Figure \ref{fig:ringvsomegab} and \ref{fig:ringvscentralden}. We suggest that the innermost and outermost $\xtwo$ orbits represent the idea of the iILR and oILR, respectively, and that they are more accurate than the linear perturbation theory for strong bars. Although knowing the location of resonances (or $\xtwo$ orbits) is insufficient to pin down the exact location of the nuclear ring, it is still informative to constrain the ring size. The exact location should be determined by the third condition discussed above.

\begin{figure}[!t]
\epsscale{1.2} \plotone{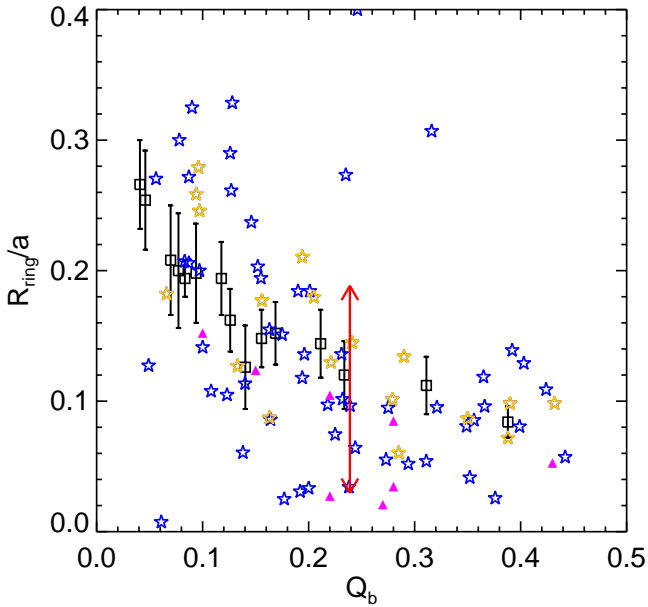}
\caption{Comparison of our numerical results with the observational measurements for the relation between the normalized ring radius $\Rring/a$ and bar strength $\Qb$. The open black squares represent the non-self-gravitating models from \citet{kim12b}, the symbols give the mean values averaged over $t=0.3$--$1.0\Gyr$, while error bars represent the standard deviations. The star symbols and filled triangles are from \citet{com10} and \citet{maz11}, respectively. Yellow stars stand for the nuclear rings in double barred galaxies from \citet{erw04} and blue stars are single barred or unbarred galaxies (for unbarred galaxies $\Qb$ is $Q_{g}$ which is defined in the same way). The $\Qb$ measured in the unbarred galaxies is defined in the same way but is mainly contributed by the strong spiral arms or oval disks (see \citealt{com10}). The red line with arrows on both ends indicates the range of the nuclear ring size in our models, with the largest ring generated from model O21C24B10 and the smallest ring generated from model O45C24B100.
\label{fig:compobs}}
\vspace{0.2cm}
\end{figure}

\subsection{An Empirical relation between $\mring$ and bar parameters}
\label{sec:empiricalrelation}

The dashed lines in Figure \ref{fig:ring_prediction} indicate the extrapolation of ring locations with their ``required'' centrifugal force barrier to form an $\xtwo$-type ring. These lines that fit our numerical results are given by
\begin{equation}\label{eq:fitomega}
f(R_{\rm ring}) = -5.23(R_{\rm ring}/\kpc)+10.46~~{\rm for}~{\rm varying}~\Omb,
\end{equation}
and
\begin{equation}\label{eq:fitcentral}
f(R_{\rm ring}) = -0.04(R_{\rm ring}/\kpc)+7.09~~{\rm for}~{\rm varying}~\rhobul.
\end{equation}

Based on the discussion above, $\mring$ correlates with the total angular momentum removal due to the bar potential, which should be a function of $\Qb$, $a/b$, and $\RCR/a$, but less sensitive to $\rhobul$ as Figure \ref{fig:ring_prediction}(b) demonstrates. The reason is that the first three parameters influence the shock formation and angular momentum loss, while $\rhobul$ just affects the angular momentum distribution in the inner part.

For an external galaxy with a bar, a nuclear ring and a well-defined rotation curve, we can simply compute $\mring$ by Equation \eqref{eq:frdefinition} as we know the location of the ring and where the bar ends. $\Qb$ or $a/b$ can be derived from the near-IR images of the galaxy (e.g. \citealt{com10}). $\Qb$, $a/b$, and $\RCR/a$ all contribute to gas angular momentum removal, which is reflected by $\mring$ (Section~\ref{sec:fomationmechanism}). If we find a correlation between $\mring$ and these bar parameters, we can put some constraints on the bar pattern speed $\RCR/a$ for real galaxies. We explore the relation between $\mring$ and $\Qb$, $a/b$, and $\RCR/a$ in a follow-up paper, and apply it to a sample of observed rings (Li, Shen \& Kim 2015, in preparation). We also suggest that since nuclear rings may not tightly correlate with the resonance radii read directly off the frequency curves, $\mring$ would be quite useful to study the properties of real galaxies that contain nuclear rings.

\subsection{Comparison with observational results}
\label{sec:compareobs}

Observations show that almost all the nuclear rings are roundish (i.e. $\xtwo$-type), and they appear to be located close to $R_{\rm to}$ (e.g., \citealt{but96,pin14}, some example galaxies are NGC1097, NGC7217, NGC5728, NGC1512, NGC1068, NGC3504, and ESO 565-11). Our numerical results are consistent with observations as most of the $\xtwo$-type rings in the simulations are within $0.5-1.0\kpc$, comparable to $R_{\rm to}$ at $\sim1.5\kpc$.

Furthermore, $R_{\rm peak}$ is where the rise of the inner rotation curve becomes shallower than $R^{0.5}$ since $f(R) \propto R^{2\alpha-1}$, and $R_{\rm to}$ is where $\alpha$ becomes $0$ as shown in Section~\ref{sec:fomationmechanism}. $R_{\rm peak}$ is closely related to, but smaller than $R_{\rm to}$\footnote{It is theoretically possible that for some bulge profiles, $f(r)$ may decrease monotonically, for which $R_{\rm peak}$ cannot be defined.}. Strictly speaking, we argue that the site of the smallest roundish ring is constrained by $R_{\rm peak}$ instead of $R_{\rm to}$, as shown in Figure~\ref{fig:ring_prediction}.

\citet{maz11} recently measured the ring radii for a sample of 13 barred/unbarred galaxies that contain star-forming nuclear rings. They argued that the ring size is anti-correlated with the compactness $\C\equiv V_0^2/R_{\rm to}$ of the galaxy, where $V_0$ is the velocity at the flat(ter) part of the rotation curve. Their results show that more compact (with smaller $R_{\rm to}$) galaxies have a smaller ring. This result seems inconsistent with our simulations at first glance. First of all, if we treat the slope of the initial steeply rising component of the rotation curve (which is proportional to the central density of the galaxy $\rho_0$) as the true compactness, $\C$ should be defined as $\C^{\prime}\equiv V_0^2/R_{\rm to}^2$ instead, but we find that it still follows a negative correlation with the ring size even with this updated definition. The range of $\C$ in our models is from $\C\sim 2.96\times10^4 \;(\rm km\;s^{-1})^2 \kpc^{-1}$ (model O33C12B10) to $\C\sim 1.08\times10^5 \;(\rm km\;s^{-1})^2 \kpc^{-1}$ (model O33C40B10), comparable to the range covered by their sample. The discrepancy is probably because their sample is well-controlled. The galaxies in \citet{maz11} exhibit a positive correlation between $\C$ and $\Qb$, as well as an obvious negative correlation between the ring thickness and $\C$, whereas $\Qb$ is nearly constant in our Figure~\ref{fig:ringvscentralden}. The anti-correlation between $\C$ and $\Rring$ in their results may simply be a reflection of a more intrinsic negative correlation between $\Rring$ and $\Qb$, as suggested by \citet{kim12b}.

Figure \ref{fig:compobs} updates the comparison of the results from \citet{com10} (star symbols) with simulation results (squares with error-bars) from \citet{kim12b} on the $\Rring/a$--$\Qb$ plane. There are two main changes in this figure compared to \citet{kim12b}. First, we divide the sample into single barred/unbarred galaxies\footnote{A possibility might be that a bar have existed in the past and is now dissolved but the nuclear rings still exist, or the nuclear ring in these unbarred galaxies might be due to strong spiral arms that could also remove angular momentum, so we consider unbarred galaxies together with single barred galaxies.} (blue stars) and double barred galaxies (yellow stars) according to the catalog in \citet{erw04}. If a secondary bar can further drive gas in the nuclear ring to the central part (e.g. \citealt{shl89}), one may then expect that the yellow stars should be systematically lower than the blue stars. However, we observe no major difference between the blue stars and yellow stars, suggesting that secondary bars may not further drive gas flow inward, similar to the results in \citet{mac02}. Second, we plot the maximum and minimum sizes of the nuclear rings formed in our simulations as shown by the red line segment with arrows on both ends. The results imply that for a fixed bar strength, the size of nuclear rings can still vary by a factor of $\sim3$, which is controlled by $\Omb$, $\rhobul$ and $\MBH$, as discussed in the present paper.

A single bar potential is unlikely to explain ultra compact nuclear rings ($\Rring/a$ less than 0.1). We suggest the magnetic field might play an important role in such compact rings since the magnetic field of equipartition strength with the thermal energy could make the ring size smaller by a factor of $\sim2$ (\citealt{ks12}). In addition, a larger effective sound speed can make the rings smaller (\citealt{kim12a}), and an extremely compact center (such as the ``BH'' we use) can also generate compact rings, like in the model O45C24B100.

\subsection{Why $\xone$-type rings are rare in Nature?}

Almost all nuclear rings that have been observed are $\xtwo$-type, which may indicate that in real galaxies the bar does not rotate too rapidly and the central density is dense enough. These two conditions are not difficult to achieve. First, there is a natural upper limit for bar pattern speed ($\RCR/a \geq 1$), as the bar supporting orbits cannot extend beyond the co-rotation radius. Since bars can interact with bulges and dark matter halos by exchanging angular momentum, the bar pattern speed can decrease with time as predicted by several N-body simulations (e.g., \citealt{deb00,ath03}). Second, the bar can also drive gas to the center, gradually resulting in an even denser center. The absence of $\xone$-type rings may be the result of a self-regulated process in Nature.

\begin{figure*}[!t]
\epsscale{1.1} \plotone{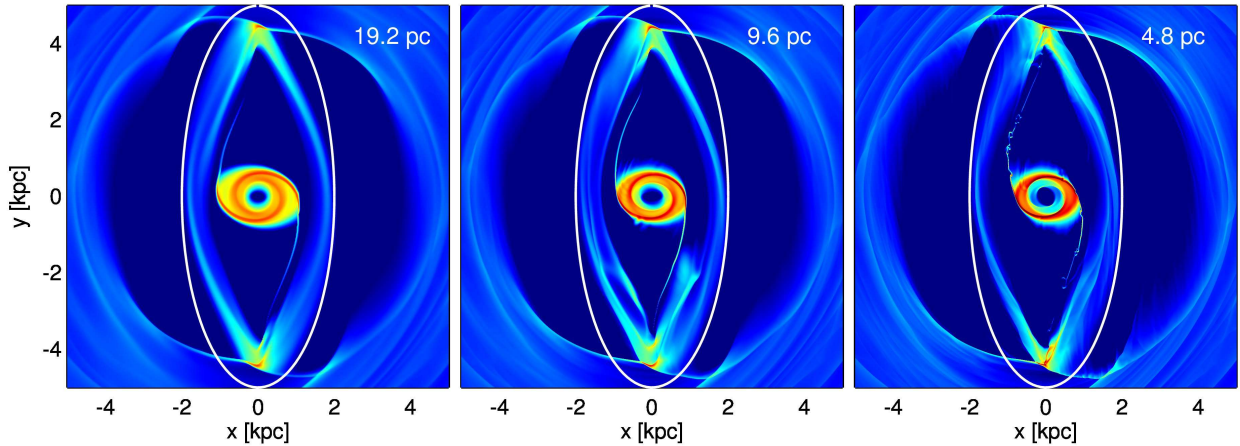}
\caption{Convergence test for the resolution effects. The three panels show the gas density at $t=800\Myr$ for the canonical model with grid spacing of $19.2\pc$, $9.6\pc$, and $4.8\pc$ from left to right, which corresponds to 0.4\%, 0.2\%, and 0.1\% of the bar size. The white ellipse denotes the bar outline. 
\label{fig:resolution}}
\vspace{0.2cm}
\end{figure*}

\subsection{Limitations of this work}
\label{sec:limitations}

In this work, we explore only the dynamical effects of $\Omb$ and $\rhobul$ on the properties of nuclear rings. The self-gravity of gas does not make a significant effect on the ring position which is the main focus of the current work, since the gravitational force due to the self-gravity in the radial direction becomes zero at the ring (or more precisely, at the radius near the middle of the ring where the potential reaches a minimum). According to \citet{kim12b}, turning on gas self-gravity may make the ring about $5\%-10\%$ larger for an initially uniform gas disk. We also double-checked that including self-gravity in our models tends to make the ring denser and clumpier, but the ring size changes little.

There are also some other unaccounted physical mechanisms that may affect the angular momentum loss of the gas and may increase the scatter of ring sizes. A strong magnetic field makes the ring rounder and smaller by about $20\%$ (\citealt{ks12}). Sound speed, star-formation and feedback may also play a role in the angular momentum transfer of gas, which could complicate the formation of nuclear rings further.

Here we use a 2D model to study the gas dynamics, but the thin-disk approximation may not be valid at small $R$ (of order of disk scale height, about a few hundred parsec). When the vertical degree of freedom is considered, the gas would exhibit motions along the vertical direction, mixing fluid elements at different heights. Since the vertical velocities, which are at sonic levels at best, would be much smaller than the in-plane velocities induced by the bar, however, finite disk thickness is not expected to change the ring positions much. Therefore, our 2D models can be viewed as the first-order approximation to what are more realistic 3D models. A similarity between 2D and 3D results can also be found by comparing the morphologies of a ring and dust lanes in 3D SPH simulations (Figure~1 of \citealt{kim11}) with those in our current 2D simulations. Note that their 3D simulations have included self-gravity, gas cooling and heating, and supernova feedback.

It is worth emphasizing that the aim in this study is not to attempt perfect realism, but rather to understand the physical mechanism of nuclear ring formation under some \textit{given}, albeit imperfect, conditions of gas.

\subsection{Resolution effects}
\label{sec:resolution}

Very recently, \citet{sor15} studied gas flow patterns under a bar potential, using a numerical setup similar to ours. Using a different grid-based code developed by \citet{van82}, they found that the size of nuclear ring formed in their simulations decreases systematically with increasing grid resolution, although their results appear not to be converged at their highest resolution (about 0.4\% of the bar size). Motivated by their results, we have made a convergence test by running two additional models with $1024\times1024$ and $2048\times2048$ cells for our fiducial set of parameters. The corresponding grid spacing is $\Delta x= 19.2\pc$ and $9.6\pc$, which are about 0.4\% and 0.2\% of the bar size. Figure \ref{fig:resolution} compares the snapshots of gas surface density at t=$800\Myr$ from the additional models as well as our standard model with $\Delta x = 4.8\pc$. Although the ring is more widely distributed in a lower-resolution model, the averaged ring radius is $620\pc$, $593\pc$, $590\pc$ for models with $\Delta x= 19.2$, $9.6$, $4.8\pc$, respectively, indicating that the nuclear ring size is almost the same for the last two resolutions. This confirms that our numerical results presented in the paper are not affected by resolution.

\section{Conclusion}

The main results of the present work are summarized as follows:

1. \emph{Nuclear rings in hydrodynamical simulations} -- Nuclear ring regions are usually composed of high-density gas and massive young stars. In our numerical simulations as gas passes through the dust-lane shocks, it loses angular momentum, flows inward, and forms a nuclear ring where the centrifugal force balances the gravity. Nuclear rings are relatively long-lived and do not vary their shape much over time, thus they should be a good indicator to study gas morphology and galaxy properties.

2. \emph{Effects of galactic parameters on nuclear ring size} -- In this paper we focus on studying the effects of varying bar pattern speed $\Omb$, bulge central density $\rhobul$, and black hole mass $\MBH$. The nuclear rings formed in our simulation can be divided into two groups: $\xtwo$-type rings which are nearly round, and $\xone$-type rings which are highly elongated along the bar major axis. We find that the size and thickness of $\xtwo$-type rings are tightly correlated with the galactic properties we varied. Galaxies with low $\Omb$ tend to form large and thick $\xtwo$-type rings, while galaxies with high $\rhobul$ form large but thin $\xtwo$-type rings. A ``BH'' could help to form $\xtwo$-type rings by removing the iILR or the innermost $\xtwo$ orbits. These correlations may be examined in future observations.

3. \emph{Nuclear ring formation mechanism} -- Observations show that nuclear rings seem to be preferably located near the turnover radius $R_{\rm to}$ of the galaxies. We suggest that the smallest $\xtwo$-type ring forms near $R_{\rm peak}$ where the normalized centrifugal force $f(R)$ reaches its local maximum. $R_{\rm peak}$ is related to, but smaller than $R_{\rm to}$. We explain the transition of an $\xtwo$-type ring to an $\xone$-type ring is due to the distribution of the centrifugal force in the inner part or the presence of the innermost $\xtwo$ orbit (or iILR). Although all $\xtwo$-type nuclear rings in our simulations seem to be located well in the range of $\xtwo$ orbits (or ILRs), knowing the resonance radii is insufficient to pin down the exact location of these nuclear rings. We suggest that an $\xtwo$-type nuclear ring forms exactly at the radius where the residual angular momentum of infalling gas balances the centrifugal force, which can be described by a parameter $\mring$ computed from rotation curves. $\mring$ can also be used to predict the nuclear ring size since it serves as a rough proxy of the total angular momentum removal of gas, and should be a function of bar strength $\Qb$, bar axis ratio $a/b$ and bar pattern speed $\RCR/a$. This will be carefully studied in a follow-up study.

\acknowledgments
We gratefully acknowledge helpful discussions with Min\ Du, Jerry\ Sellwood, Ortwin\ Gerhard, Kartik\ Sheth, and Sarah\ Bird. We thank Yonghwi\ Kim for helping with the \textit{Athena} code. Hospitality at APCTP during the 7th Korean Astrophysics Workshop is also kindly acknowledged. The research presented here is partially supported by the 973 Program of China under grant no. 2014CB845700, by the National Natural Science Foundation of China under grant nos.11333003, 11322326, 11073037, and by the Strategic Priority Research Program  ``The Emergence of Cosmological Structures'' (no. XDB09000000) of the Chinese Academy of Sciences. This work made use of the facilities of the Center for High Performance Computing at Shanghai Astronomical Observatory. The work of W.-T. Kim was supported by the National Research Foundation of Korea (NRF) grant, No. 2008-0060544, funded by the Korea government (MSIP).

\end{document}